\documentclass[modern]{aastex61}

 \usepackage{natbib}
\usepackage{lineno}
 \usepackage{graphicx}

\newcommand{\etal}{{\it et al.}}
\newcommand{\ie}{{\it i.e.}~}
\newcommand{\eg}{{\it e.g.},~}

\newcommand{\Rmnum}[1]{{\uppercase\expandafter{\romannumeral #1}}}
\usepackage{amsmath}
\usepackage{longtable}
\usepackage[colorinlistoftodos]{todonotes}
\usepackage{hyperref}

\newcommand{\rsun}{$R_{\odot}$}
\hypersetup{
    colorlinks=true,       
     linkcolor=blue,          
     citecolor=blue,        
     filecolor=black,      
      urlcolor=blue           
}

\shorttitle{CME Kinematics and Magnetic Reconnection}
\shortauthors{Zhu \etal}

\begin{document}

\title{How Does Magnetic Reconnection Drive the Early Stage Evolution of Coronal Mass Ejections?}

\correspondingauthor{Chunming Zhu}
\email{chunming.zhu@montana.edu}

\author{Chunming Zhu}
\affiliation{Physics Department, Montana State University, Bozeman, MT 59717, USA}

\author{Jiong Qiu}
\affiliation{Physics Department, Montana State University, Bozeman, MT 59717, USA}

\author{Paulett Liewer}
\affiliation{Jet Propulsion Laboratory, California Institute of Technology, Pasadena, CA 91109, USA}

\author{Angelos Vourlidas}
\affiliation{Johns Hopkins University Applied Physics Laboratory, Laurel, MD 20723, USA}
\affiliation{IAASARS, National Observatory of Athens, GR-15236, Penteli, Greece}

\author{Michael Spiegel}
\affiliation{Physics Department, University of California, Davis, CA 95616, USA}

\author{Qiang Hu}
\affiliation{Department of Space Science/CSPAR, University of Alabama in Huntsville, Huntsville, AL 35805, USA}

\begin{abstract}
    Theoretically, CME kinematics are related to magnetic reconnection processes in the solar corona. However, the current quantitative understanding of this relationship is based on the analysis of only a handful of events. Here we report a statistical study of 60 CME-flare events from August 2010 to December 2013. We investigate kinematic properties of CMEs and magnetic reconnection in the low corona during the early phase of the eruptions, by combining limb observations from STEREO with simultaneous on-disk views from SDO. For a subset of 42 events with reconnection rate evaluated by the magnetic fluxes swept by the flare ribbons on the solar disk observed from SDO, we find a strong correlation between the peak CME acceleration and the peak reconnection rate. Also, the maximum velocities of relatively fast CMEs ($\gtrsim 600$ km~s$^{-1}$) are positively correlated with the reconnection flux, but no such correlation is found for slow CMEs. A time-lagged correlation analysis suggests that the distribution of the time lag of CME acceleration relative to reconnection rate exhibits three peaks, approximately 10 minutes apart, and on average, acceleration-lead events have smaller reconnection rates. We further compare the CME total mechanical energy with the estimated energy in the current sheet. The comparison suggests that, for small-flare events, reconnection in the current sheet alone is insufficient to fuel CMEs. Results from this study suggest that flare reconnection may dominate the acceleration of fast CMEs, but for events of slow CMEs and weak reconnection, other mechanisms may be more important.
    
\end{abstract}

\section{Introduction}
Coronal Mass Ejections (CMEs), originally observed in white light by the coronagraph as discrete and bright features moving outward from the Sun (\citealp{tousey1973solar,gosling1974mass}), have also been detected in extreme ultra-violet (EUV), soft X-rays, and radio (reviews by \citealp{alexander2006brief,hudson2006coronal}). Typically, the CMEs carry masses ranging from $10^{13}$ to $10^{16}$ g, with velocities between several tens to a few thousand km~s$^{-1}$, and energies of $10^{22-25}$ J (\citealp{vourlidas2002mass,yashiro2004catalog,vourlidas2010comprehensive}). A comparison of the coronal energy sources and the required energy of typical CMEs suggests they are generally driven by the release of magnetic energy (\citealp{forbes2000review}).

MHD models have been developed to study the evolution of the coronal magnetic field leading to the CME eruption (see recent reviews by \citealp{chen2017physics, green2018origin}). In these models, the coronal magnetic structure rises slowly due to gradually changing boundary conditions in the photosphere, such as a shear motion, converging motion, or flux emergence. Free magnetic energy in the corona is stored during this quasi-equilibrium evolution. At a certain point in this evolution, an abrupt eruption can be triggered either by an ideal MHD instability or loss of equilibrium (\eg \citealp{Lin2000, fan2007onset,torok2004ideal, kliem2006torus, aulanier2010formation}), or magnetic reconnection (\citealp{mikic1994disruption,Antiochos1999}). The onset of magnetic reconnection may also restructure the previously stable magnetic configuration by adding magnetic twist to a pre-existing flux rope, or by forming a twisted flux rope from a sheared arcade (\citealp{van1989formation,mikic1994disruption,Longcope2007a,priest2017flux}), which then erupts.  During the eruption, a CME can reach speeds of several hundred km~s$^{-1}$ within a few minutes (\eg \citealp{patsourakos2010genesis}). 

Although magnetic reconnection in the corona is a crucial process, it is not currently directly observable. However, magnetic reconnection forms closed loops and energizes plasmas and particles therein, resulting in impulsively enhanced flare radiation. 
The observed coincidence  between the evolution of the CME kinematics and that of the flare X-ray fluxes indicates that the CME eruption is related with the reconnection which produces solar flares (\citealp{zhang2001temporal,zhang2004study,temmer2008acceleration,bein2012impulsive,patsourakos2013direct}), as predicted by CME initiation models (\eg \citealp{chen2011coronal}). Apart from flare light curves, morphological evolution of flare emission in the lower atmosphere has also been used to infer the reconnection process in the corona and indirectly measure the rate of magnetic reconnection \citep{forbes1984numerical, poletto1986macroscopic}. With such measurements,  \citet{qiu2004magnetic} analyzed two well-observed CME-flare events and demonstrated that the acceleration of the studied CMEs was temporally correlated with the magnetic reconnection rate estimated from the magnetic flux swept through by flare ribbons (\citealp{qiu2002motion}). This correlation is also shown in a few recent case studies of \citet{hu2014structures} and \citet{song2018acceleration} using observations with higher tempo-spatial resolutions than earlier studies.
Because of the difficulty in observing CME evolution and measuring CME acceleration in the low corona, many studies instead compare the CME velocity
measured at a few solar radii after the peak acceleration phase, and the total magnetic flux 
reconnected during the flare. These studies found that, in fast CME events, these two quantities are positively correlated (\citealp{qiu2005magnetic,deng2017roles,toriumi2017magnetic,tschernitz2018reconnection,pal2018dependence}). Note that most of these studies have focused on fast CMEs and strong flares, and it is unclear whether this relationship applies to a broader range of CMEs, in particular, slow CMEs. 

Magnetic reconnection almost always occurs during a CME eruption, yet it is not clear whether reconnection initiates the eruption or it is a consequence of the eruption.
Investigating the relative timing between CME acceleration and flare reconnection over a large number/types of events should shed light on this question (\eg \citealp{Karpen2012mechanisms}). For such a study, it is essential to simultaneously measure CME acceleration and flare reconnection signatures in the low corona during the early phase of the eruption. Between 2010 and 2013, the  separation between the STEREO and SDO spacecraft (80 -- 140 degrees) provides a unique opportunity to observe CMEs from the limb and associated flare signatures and magnetic field evolution in the source regions on the solar disk. These missions have obtained observations with unprecedented temporal cadence and spatial resolution, crucial for quantifying the dynamics in the early phase of the eruption. Here, we carry such a statistical study of a large sample of CME-flare events, using  observations from multiple viewpoints: we track the CME evolution in the low corona via limb observation by STEREO, and measure the magnetic reconnection rate by analyzing flare signatures and magnetograms in  simultaneous on-disk observations by SDO.

 In Section 2, we provide the related information about SDO and STEREO, and the methods used here for the measurements of the CME kinematics and reconnection properties during the early stage evolution of the CME-flare events. The statistical results are presented in Section 3 on the CME kinematics and the reconnection process, and in Section 4 on the time evolution of the CME kinematics, flare reconnection, and flare X-ray emission. We then evaluate energetics for these events in Section 5.  Conclusions and discussion are given in Section 6.

\section{Data and method}
\subsection{Data}
The data set under investigation start from the launch of SDO (\citealp{pesnell2012solar}) in 2010, continuing until late 2013. During this period, the separation angle between STEREO (\citealp{kaiser2008stereo}) and the Earth (or SDO) increased from $\sim$80 to $\sim$140 degrees. STEREO is composed of twin spacecraft: STEREO-A (Ahead) and STEREO-B (Behind). Three instruments from the Sun Earth Connection Coronal and Heliospheric Investigation (SECCHI; \citealp{howard2008sun}) on board each spacecraft have been used to track the early evolution of the Earth-side CMEs which are located near the solar limb when viewed from STEREO. These three instruments include the Extreme Ultraviolet Imager (EUVI;  $<$1.7 \rsun; \citealp{wulser2004euvi}), the inner coronagraph COR1 (1.3 -- 4 \rsun; \citealp{thompson2003cor1}) and the outer coronagraph COR2 (2.5 -- 15 \rsun). For EUVI, we prefer to use high-cadence (75 s) images in 171~{\AA} when there are such data available; otherwise we use 195~{\AA} observations at $\sim$5 min cadence. For each event observed by STEREO, a single viewpoint is chosen (Table~1) based on a closer proximity of the source region to the Earth-side limb and a higher data cadence if it exists. 

The on-disk observations are provided by SDO spacecraft. The Atmospheric Imaging Assembly (AIA; \citealp{lemen2011atmospheric}) on board SDO takes full-disk images of the Sun in 10 EUV/UV channels ($\log{T}$ $\sim$ 3.7 -- 7.3), with $\sim$12 -- 24 s cadence. Full-disk magnetograms are from the Helioseismic and Magnetic Imager (HMI; \citealp{schou2012design}), with 1$''$ spatial resolution and 45 s cadence. When choosing the CME-flare events, we require that a flare is associated with each CME by checking if flare ribbons can be clearly detected in UV (using 1600 {\AA} on transition region and upper photosphere) and/or EUV (here using 304 {\AA} on transition region and chromosphere). Then we determine the CME source regions based on the timing and location of the flares. 

Examining joint STEREO and SDO observations between 2010 and 2013, we have selected 60 CME-flare events, listed in Table~1. For these events, STEREO observations have very well covered the initial stage of the CME. 54 out of the 60 CME-flare events are located within $\pm$30 degrees from the limb in the STEREO field of view. 
The source regions of the 60 CMEs are identified from SDO observations, and we find that 27 events originate from active regions (ARs), and the other 33 from quiet regions or decaying plage regions with weak magnetic fields.
Using GOES light curves, we have further identified magnitudes of flares associated with 41 of the CMEs, including 1 X, 11 M, 16 C, and 13 B class flares. For the remaining 19 events the GOES X-ray curves are either affected by the background emission or other flaring regions on the solar disk, or have no flare information available. We track the time evolution of the CME height and flare ribbons in the source region, and measure and compare CME kinematics and flare reconnection properties in these events.

\subsection{Measurement of CME Kinematics}
\begin{figure*}[!ht]
  \begin{center}
        \includegraphics[viewport = 43 389 511 613, clip, width=0.95\textwidth]{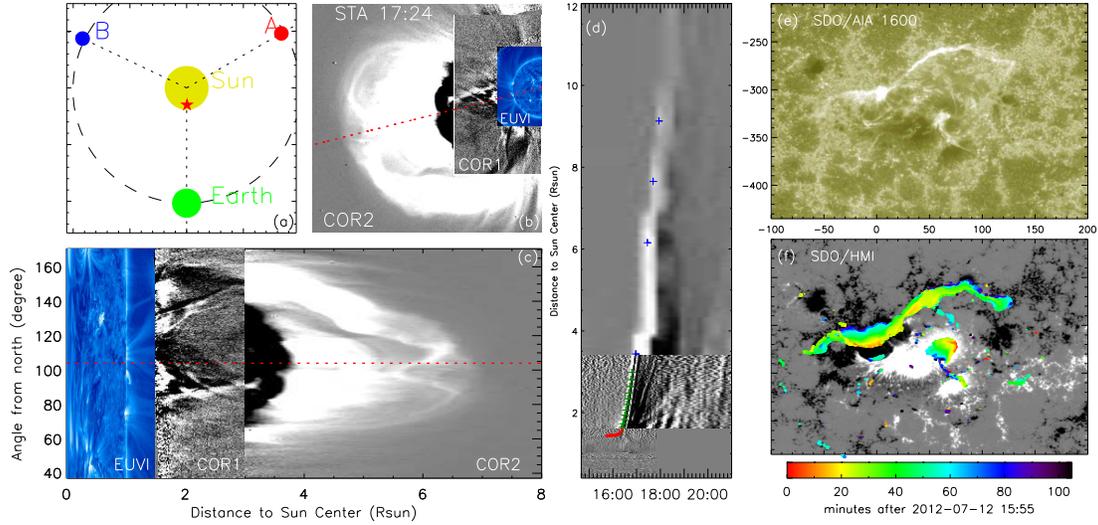}
        
    \caption{STEREO and SDO observations of a CME on 2012/07/12 (SOL2012-07-12T16:53), illustrating the procedures used in this study for deriving a CME's early trajectory and the associated magnetic reconnection rate. (a) Positions of STEREO/A\&B, and SDO around the Sun. The CME source region is indicated by the star sign. (b) Joint observations from STEREO/A EUVI 171 \AA, COR1, and COR2. (c) The Plate-Carr\'{e}e projection of the joint observations. A slit (dashed line) along erupting features is chosen to generate a time-space stackplot (J-map). The slit location in the original images is also shown in (b). (d) The generated J-map, with trajectories of the tracking features indicated by the plus signs. (e) A snapshot of the flaring ribbons, from SDO/AIA 1600 \AA. (f) The expanding flare ribbons in 1600 {\AA} (colored) which swept the magnetic fields measured by SDO/HMI (gray).}
    
    \label{fig:f1}
  \end{center}
\end{figure*}

An example of a CME-flare observed on 2012/07/12 (SOL2012-07-12T16:53) is displayed in Fig.~\ref{fig:f1}, which demonstrates the methods we used to track a CME and measure the reconnection flux. To help identify the structures or features in solar images, we use the wavelet processed images for EUVI\footnote{http://sd-www.jhuapl.edu/secchi/wavelets/} (\citealp{stenborg2008fresh}), while the COR1 and COR2 data are first processed with secchi\_prep and then converted to base differences by subtracting a pre-eruption image, see Fig.~\ref{fig:f1}(b). Then the images are projected to a rectangular Plate Carr\'{e}e (PC) system (\citealp{thompson2006coordinate}) with axes of radial distance from solar disk center and position angle measured counter-clockwise around the Sun from solar north (Fig.~\ref{fig:f1}(c)). A virtual slit with a width of 2 degrees and along the center of the erupting CME in the PC-projected images is chosen to generate time-height stackplots for each of the three instruments. The stackplots are then co-aligned, to form a  map displaying the CME trajectory, \ie  a ``J-map'' (\citealp{sheeley1999continuous}), as shown in Fig.~\ref{fig:f1}(d). We note that apparent radial CME eruptions have been preferred during our data selection. The CME front is tracked in  COR1 and COR2. The CME front is sometimes difficult to identify in EUVI images due to the temperature and density response of the EUV bandpasses and/or the reduced compression at the front in the early evolution of the eruption. Therefore, we manually select the foremost detectable feature in the EUV, recognizing that it may not always correspond to the white light front. As shown in Fig.~\ref{fig:f1}(d), the identified heights of the feature(s) are superimposed on the J-map. We estimate the errors of the heights in EUVI, COR1, and COR2 to be 4, 2, and 2 pixels, respectively (\eg \citealp{song2018acceleration}).

\begin{figure*}[!ht]
  \begin{center}
        \includegraphics[viewport = 95 257 441 755, clip, width=0.8\textwidth]{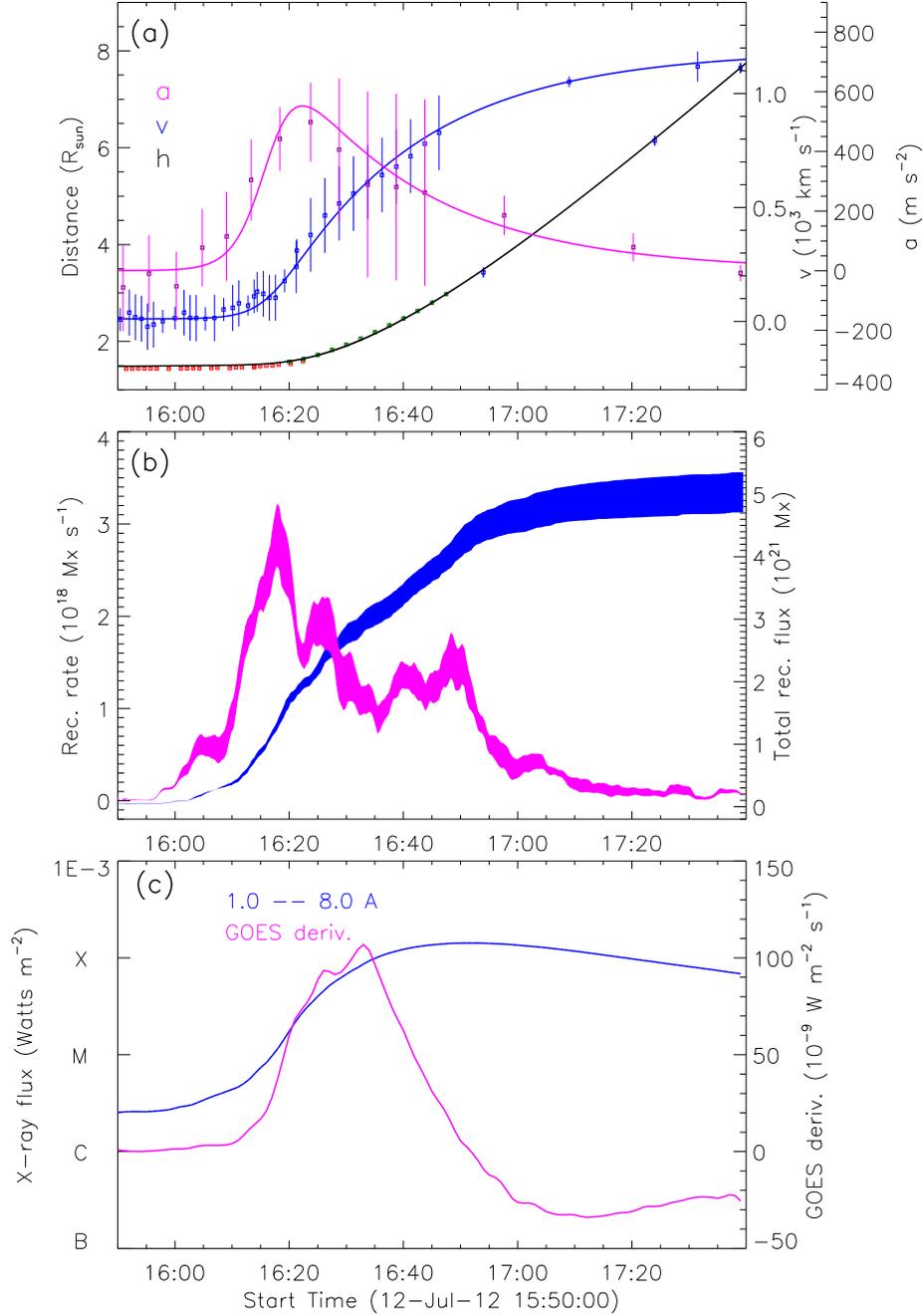}
        
    \caption{CME kinematics and a related X1.4 flare of SOL2012-07-12T16:53. (a) Projected heliocentric height ($h$), velocity ($v$), acceleration ($a$) of the CME. For $h$, the red, green, and blue squares delineate the trajectories from STEREO/A EUVI, COR1 and COR2, respectively. The blue curve of $v$ corresponds to the fit. The other two curves (black and pink) are from the integration and derivative of the fitting of $v$, respectively. (b) Magnetic reconnection rate and the accumulated reconnection flux. (c) GOES X-ray flux of 1.0--8.0~{\AA} and its time derivative.}
    
    \label{fig:f2}
  \end{center}
\end{figure*}

With the CME height-time measured, it is straightforward to compute the projected CME velocity and acceleration by using the time derivatives (Fig.~\ref{fig:f2}(a)). For SOL2012-07-12T16:53, its acceleration increased abruptly from nearly zero at 16:05 UT to a peaking value of ${\sim}500~m~s^{-2}$ at 16:23 UT, then gradually decreased to ${\sim}10~m~s^{-2}$ in 1.3 hours. This acceleration profile is similar to that of the CME reported by \citet{gallagher2003rapid}, both suggesting an exponential rise followed by another exponential decay. Here we follow \citet{gallagher2003rapid} and model the CME acceleration with the two-phase exponential function in the form of $a(t)=[1/[a_r~exp(t/\tau_r)]+1/[a_d~exp(-t/\tau_d)]]^{-1}$, where $a_r$ and $a_d$ are the magnitudes of acceleration in the two phases, and $\tau_r$ and $\tau_d$ are the corresponding e-folding times, respectively. These four parameters characterize CME kinematics in the early phase.

To derive the four parameters in the acceleration function, we fit the observed CME velocity profile, obtained as the time derivative of the CME height measured from EUVI, COR1, and COR2 observations, to the time integral of the acceleration function $v(t)=v_0+\int a(t)dt$, where $v_0$ is the fifth parameter from the fitting. We choose to fit the velocity profile because it exhibits a relatively smooth-varying behavior. On the other hand, there can be a gap in the CME heights measured between EUVI and COR1, and between COR1 and COR2, most likely due to different sensitivities to various parts of a CME by different instruments observing in different spectral ranges. The uncertainty in tracking the weak signatures of the CME leading edge, together with the varying time cadence, also results in spurious features in the CME acceleration profile derived as the second time derivative of the measured CME height. Therefore, fitting to the velocity profile is an optimal choice for modeling CME acceleration.

 Fig.~\ref{fig:f2}(a) shows the model CME velocity profile $v(t)$ from the fitting (the blue/middle curve) superimposed with the measured CME velocity (blue symbols,  smoothed with a boxcar of three measurements) for the event SOL2012-07-12T16:53 (Fig.~\ref{fig:f1}). The acceleration profile $a(t)$, which is calculated as the time derivative of $v(t)$, is presented as the pink/top curve; it agrees well with observational measurements. Finally, the time integral of $v(t)$, $h(t)=h_0+\int v(t)dt$, is fitted to the heights of the CME front measured in COR1 and COR2 observations (EUVI observations are not included here because of possible inconsistency of the tracked features) to yield the last parameter, the CME initial height $h_0$. Here $h_0$ corresponds to an upper limit if considering possible pileup at the CME front (\eg \citealp{howard_evolution_2018}). For the SOL2012-07-12T16:53 event, we find $v_{0}$=11.4~km~s$^{-1}$, and $h_{0}=0.47$ \rsun. Note that $h(t), v(t)$, and $a(t)$ refer to the CME height, velocity, and acceleration projected on the plane of the sky (POS), respectively.  To compensate for the projection effect, a factor of 1/$\cos{\theta}$ is multiplied to each projected height, where $\theta$ is the longitudinal separation angle between the source region viewed from SDO and the limb of the observer (STEREO A or B) by simply assuming that the eruptions are radial. Table 1 lists the parameters characterizing CME kinematics for all the 60 CME events.
 
\subsection{Measurement of Reconnection Properties}
We also analyze flare signatures in CME source regions to infer the rate of magnetic reconnection using SDO observations. Magnetic reconnection occurring in the corona forms flare loops, and releases energy in these loops; accelerated particles or thermal conduction travel along newly reconnected field lines (flare loops) and deposit energy at their footpoints in the chromosphere, causing the brightenings there thus forming the flare ribbons in optical and ultraviolet wavelengths (\citealp{forbes1984numerical,poletto1986macroscopic, longcope2015gas}). Therefore, magnetic reconnection, which occurs in the corona yet can be hardly measured there, is mapped in the lower atmosphere by flare radiation, and the reconnection rate is measured by summing up photospheric magnetic fluxes covered by spreading flare ribbons. This method has been widely used to measure the magnetic reconnection rate (\eg \citealp{qiu2002motion,qiu2004magnetic,isobe2002reconnection,Asai2004flare, Krucker2005hard, Miklenic2007reconnection,veronig2015magnetic,kazachenko2017database,zhu2018two}). In this study, the brightening pixels are collected from AIA imaging observations in 1600 {\AA} or 304 {\AA} passband to measure the reconnection rate in flares ranging from X-class to B-class.

Flare emission in the 1600 {\AA} passband is primarily produced by enhanced C~{\sc iv} line emission due to heating at the chromosphere/transition region foot-points of reconnection formed flare loops. The flaring pixels are chosen automatically with the criteria that the intensity (in the unit of data number) is greater than $\sim$4 times the median intensity of the region of interest before the flare, and that the brightening should persist for at least 3 minutes. Such threshold is chosen to be above the mean intensity of the active region plage \citep{qiu2010reconnection}, and the persistence requirement effectively removes spurious brightening features such as short-lasting brightening due to projection of erupting filaments. Instantaneous reconnection flux is measured by summing up HMI-measured longitudinal magnetic flux swept by newly brightened flare ribbon pixels. The events selected for this study were observed near the disk center from the SDO point of view. In this study, we do not correct the projection effect, and do not extrapolate the photosphere magnetic field to the chromosphere where flare ribbons form, considering that the two effects partially cancel each other. Uncertainties in the measured reconnection flux are derived by varying the intensity threshold between 3 and 5 times the pre-flare median intensity, and also by the difference in the measured magnetic flux in the positive and negative polarities. 

For weak flares, usually below C class, emission in the 1600 {\AA} passband is not significantly enhanced against the background; alternatively, we use observations in the 304~{\AA} passband to select flare pixels with an intensity threshold of 8 times the pre-flare median intensity. The 304~{\AA} channel is dominated by He~{\sc ii} line. It is more sensitive to weak heating signatures and can help capture flare ribbons in minor flares below C-class. The difference in the measured reconnection flux using these two passbands for the same flares of C to low M-class is about 20\%, which is usually within the uncertainty of the $\Phi_{r}$ measurement. In large flares, significantly enhanced chromosphere emission tends to saturate the 304~{\AA} passband, and this passband also captures emission in post-flare loops during the late phase of the flare; therefore, we do not use 304~{\AA} observations to measure reconnection flux in large flares. 

The example of the SOL2012-07-12T16:53 showing the flaring ribbons (Fig.~\ref{fig:f1}(e)) sweeping through the magnetic fields is given in Fig.~\ref{fig:f1}(f), where the colors indicate the flux reconnected in different time intervals (see colorbar). Among the 60 CME-flare events in our database, we analyze flare ribbon evolution and measure the reconnection fluxes using 1600 (304)~\AA\ passband in 26 (16) events, respectively, while in the remainder 18 events of weak flares, reconnection flux is not measured, because contamination of emissions by erupting filaments or post-flare loops is difficult to be separated from flare ribbon emissions. For the 42 flares, the measured total reconnection flux $\Phi_{r}$ is in the range $1.5\times 10^{20}$ -- $5.5 \times 10^{21}$ Mx, with an average uncertainty of 35\%. The rate of flare reconnection is the time derivative of the reconnection flux, in units of Mx s$^{-1}$. The total reconnection flux and peak reconnection rate for the 42 flares are listed in Table~1.

\begin{figure*}[ht]
  \begin{center}
        \includegraphics[viewport =  29 329 374 613, clip, width=0.6\textwidth]{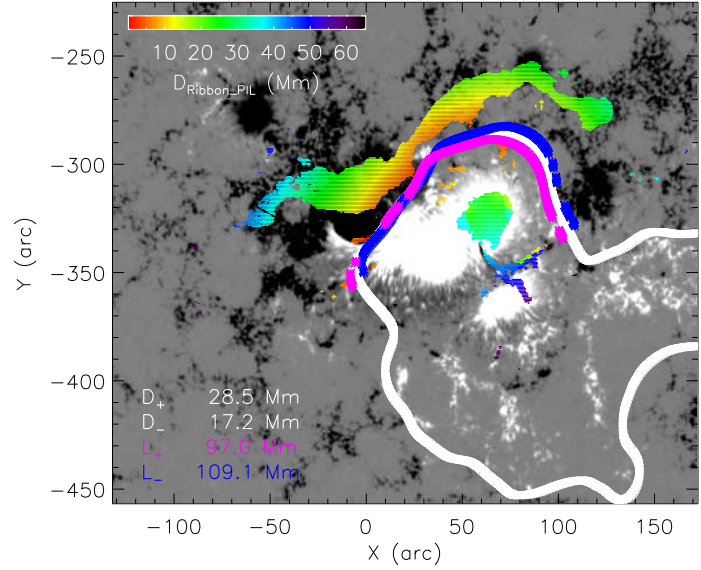}
    \caption{Measurement of the flare ribbon geometry for SOL2012-07-12T16:53 (Fig.~\ref{fig:f1}{\&}\ref{fig:f2}). The ribbon lengths $L_+$ ($L_-$) along the PIL (white thick curve) are measured by projecting the flare ribbon located in the positive (negative) polarity to the PIL, respectively. Similarly, the ribbon distances of $D_+$ and $D_-$ in the opposite polarities are determined by averaging the distances of the flaring pixels to the PIL. The color scale along the ribbons denotes the PIL distance. The HMI LOS magnetogram is shown as gray background.}
    
    \label{fig:f_rg}
  \end{center}
\end{figure*}

With the flare ribbon pixels identified in the 42 events, we also evaluate two important properties, \ie the length of the ribbons along the interface of the positive and negative magnetic polarities (polarity inversion line; PIL) $L_{R}$, and the average distance of the flare ribbons to the PIL $D$. Same example of SOL2012-07-12T16:53 is displayed in Fig.~\ref{fig:f_rg}. The ribbon pixels in the positive and negative polarities are perpendicularly projected to the PIL respectively, thus giving the corresponding lengths along the PIL, $L_{+}$ and $L_{-}$. The ribbon length is the average of $L_{+}$ and $L_{-}$, \ie $L_{R}=(L_{+}+L_{-})/2$. The average perpendicular distances of the ribbon pixels to the PIL is the ribbon distance $D$. For this flare, $L_{R}$ and $D$ have values of 103 Mm and 23 Mm, respectively. Measurements of these geometric properties are used to estimate energetics in the reconnecting current sheet in the corona (Section 5).

\section{Statistics of CME kinematics and flare reconnection properties}
\subsection{CME kinematics}
\begin{figure*}[!ht]
  \begin{center}
        \includegraphics[viewport = 57 145 717 477, clip, width=0.95\textwidth]{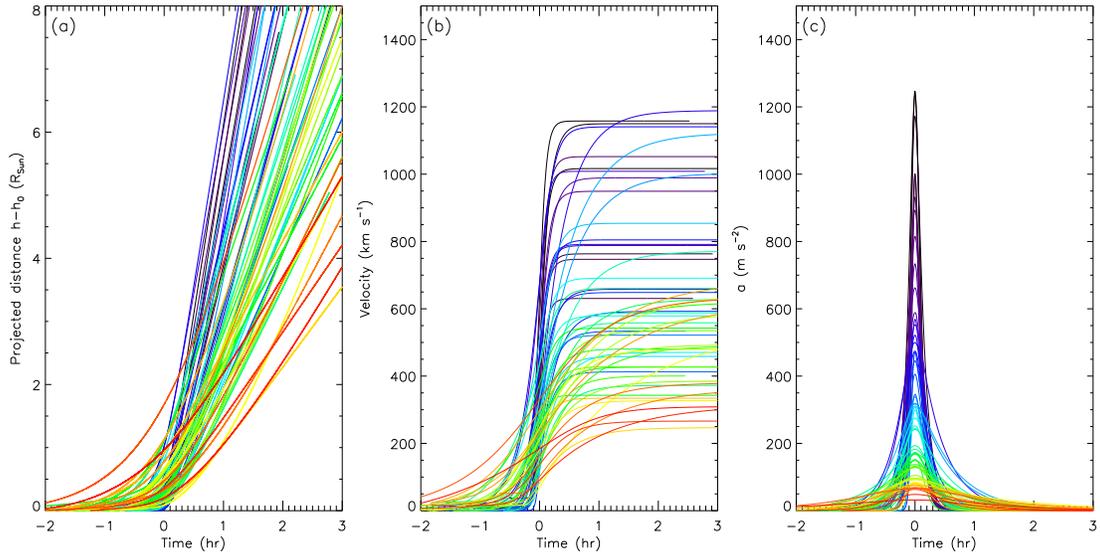}
    \caption{Overview of the kinematics of the selected 60 CMEs, including heights (a), velocities (b), and accelerations (c). The curves are co-aligned relative to the time of peak acceleration for each event. The colors of the curves are based on the peak acceleration values, as seen in (c).}
    
    \label{fig:f_cv}
  \end{center}
\end{figure*}

The selected 60 CME-flare events were observed from August 2010 to December 2013, during the rise of solar cycle 24. Following the methods introduced in Section~2, we track the CME eruptions from the solar limb to 15 \rsun, covered by EUVI, COR1 and COR2 on board STEREO. Fig.~\ref{fig:f_cv} shows the fitted profiles (heights, velocities, and accelerations projected on the POS) of the 60 CMEs. For comparison, we aligned the time profiles for each event, so that the zero time is at the peak acceleration of that event. These plots reveal a clear trend that fast CMEs are accelerated for shorter durations yet with greater acceleration rates in the \textit{low corona}, as shown by \citet{zhang2006statistical}. All the kinematic profiles (along with the error bars) of the 60 CME-flare events can be found online at MSU website\footnote{http://solar.physics.montana.edu/czhu/outgoing/CF60/}.

\begin{figure*}[!ht]
  \begin{center}
        \includegraphics[viewport = 46 418 722 613, clip, width=0.95\textwidth]{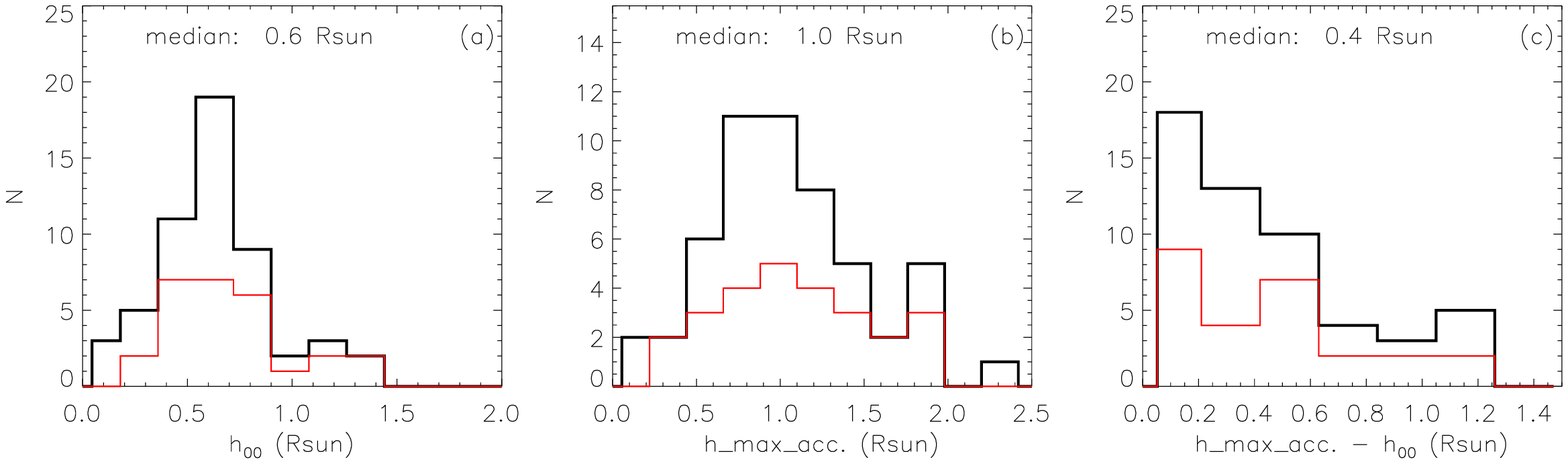}
        \includegraphics[viewport = 46 418 722 613, clip, width=0.95\textwidth]{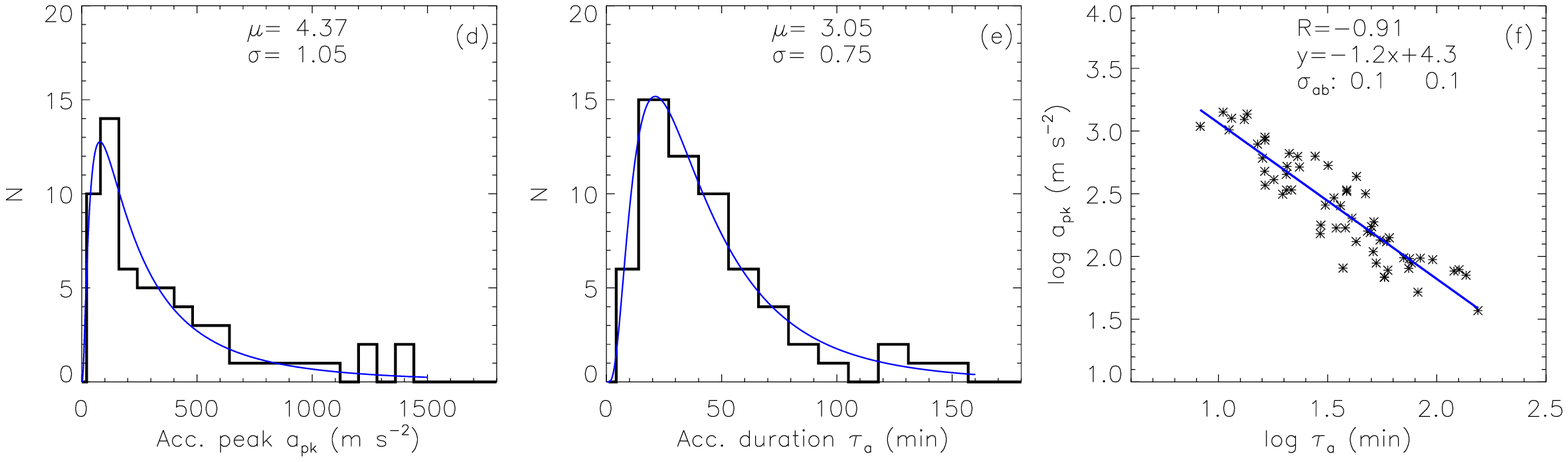}   
    \caption{Histograms of the CME heights, acceleration peaks and durations. (a) Initial heights ($h_{00}$) of the CME fronts above solar surface.  (b) The heights of each CME at maximum acceleration. (c) The distances between the $h_{00}$ and the height of of CME front at the maximum acceleration. (d) The peak values of CME accelerations. (e)  The duration of CME acceleration. (f) Relationship of the peak and duration of CME acceleration. The histograms for all events are indicated with black curves, while those for AR CMEs are indicated with red curves. The blue curves in (d) and (e) give the log-normal fittings, with the parameters noted at top. A linear fitting in (f) is included, with the correlation coefficient R, the correlation, and the uncertainties $\sigma_{ab}$ of the corresponding parameters displayed at the top.
} 
    
    \label{fig:f_kin}
  \end{center}
\end{figure*}

With these plots, we study the initial height of the CMEs, and for how far CMEs are accelerated. Histograms of the de-projected heights of the CME fronts are given in Fig.~\ref{fig:f_kin}(a--c). The initial heights of the CME fronts ($h_{00} \equiv h_0/\cos\theta$) primarily range from 0.4 -- 1.0 {\rsun} above the solar surface, with a median value of 0.6 {\rsun} (Fig.~\ref{fig:f_kin}(a)). Both the value of $h_0$ (an upper limit of the initial height) and the projection correction could have led to slightly larger values than those in \citet{bein2011impulsive}. For most of the CMEs the acceleration peaks at heights ($h_{max\_acc}$) between 0.5 and 1.9 \rsun, with a median value of 1.0 {\rsun} (Fig.~\ref{fig:f_kin}(b)). The distance between the two heights of $h_{00}$ and  $h_{max\_acc}$ ranges from 0.15 to 1.2 \rsun, with a median value of 0.4 {\rsun} (Fig.~\ref{fig:f_kin}(c)). The red curves in Fig.~\ref{fig:f_kin} are  events originating from active regions (AR, see Table 1), which exhibit very similar height (distance) distributions. 

The distributions of the acceleration properties from the fitted curves (Fig.~\ref{fig:f_cv}(c)) are shown in Fig.~\ref{fig:f_kin}(d--f). The acceleration duration here is measured using the {\em full width at half maximum} of the acceleration profiles. The peak magnitude and duration of the CME accelerations tend to follow log-normal distributions 
(Fig.~\ref{fig:f_kin}(d\&e)), similar to \citet{bein2011impulsive}. The peak acceleration $a_{pk}$ and duration of the acceleration $\tau_a$ are anti-correlated with a Pearson correlation coefficient R = -0.91, and their relationship can be described by a power-law $a_{pk} = 10^{4.3} \tau_a^{-1.2}$ (Fig.~\ref{fig:f_kin}(f)). These results are consistent with previous studies ( \citealp{zhang2006statistical,vrvsnak2007acceleration,bein2011impulsive}). 

We note that \citet{zhang2006statistical} analyzed 50 CMEs observed by LASCO from 1996 June to 1998 June, at the start of solar cycle 23, and \cite{bein2011impulsive} analyzed 90 CMEs observed by STEREO between 2007 January and 2010 May during the transition between solar cycle 23 and 24. The 60 CMEs in this study were observed by STEREO from 2010 August to 2013 December, from the rise to maximum of the solar cycle 24, and the kinematic properties of CMEs in this study are derived using a method different from \citet{zhang2006statistical} and \citet{bein2011impulsive}. The consistency of the results in these different studies, therefore, suggests that the statistical properties of CME kinematics, illustrated in Fig.~\ref{fig:f_kin}, do not depend on the phase of the solar cycle.

\subsection{Comparison between CME kinematics and reconnection properties}
\begin{figure*}[ht]
  \begin{center}
        \includegraphics[viewport = 46 427 725 613, clip, width=0.95\textwidth]{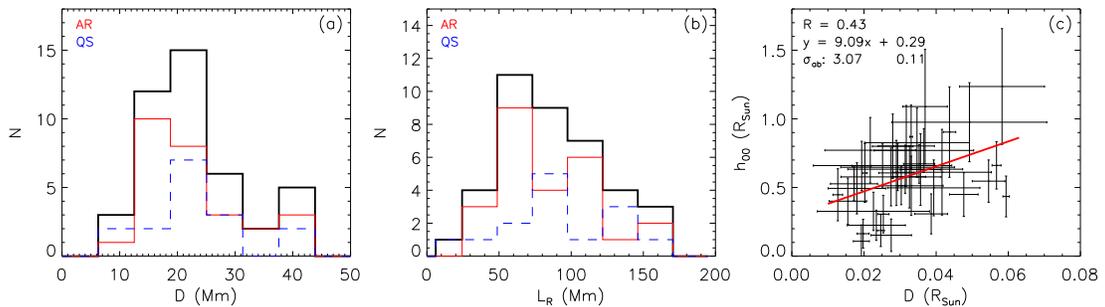}
        
    \caption{Distributions of the average distances of the flare ribbons to the PIL $D$ for the CME-flare events (a), the ribbon length along PIL (b), the correlation between $D$ and $h_{00}$ (c).}
    
    \label{fig:f_fs}
  \end{center}
\end{figure*}

Out of the 60 CMEs, 42 have reconnection properties unambiguously measured using flare ribbon signatures.  Fig.~\ref{fig:f_fs}(a\&b) display the histograms of the mean perpendicular distance $D$ from the ribbon to the PIL and the mean flare ribbon length $L_R$ along the PIL. $D$ ranges from 8 to 40 Mm, with a median value of 20 Mm in AR, slightly smaller than 23 Mm in the quiet Sun (QS) regions. The distributions of $L_{R}$ are similar between AR and QS, both centering at around 80 Mm.  Interestingly, when comparing $D$ with the CME initial heights $h_{00}$ (Fig.~\ref{fig:f_fs}(c)), though  $D$ is small compared to $h_{00}$ ($\sim$5--20\%), there is a trend (R = 0.43) that larger ribbon separations are associated with larger $h_{00}$. Flare ribbons are produced by energy release due to reconnection in the corona, and in general, reconnection occurring at a higher altitude forms larger flare loops with a greater ribbon separation. Therefore, the $h_{00}$ -- $D$ plot in the figure suggests a scaling relationship between the initial height of the CME front and the height of flare reconnection, and these two parameters provide the characteristic scale of the flux system involved in the eruption.

\begin{figure*}[ht]
  \begin{center}
        \includegraphics[viewport = 120 329 643 613, clip, width=0.95\textwidth]{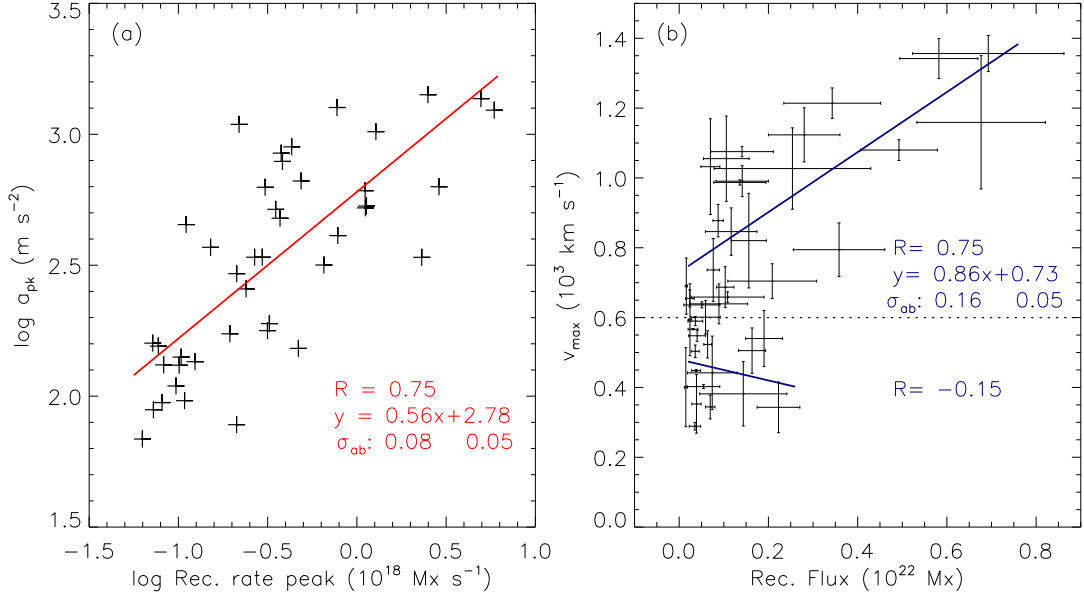}
    \caption{Relationship between the peaks of the acceleration and reconnection rate (a), and between the CME maximum velocities and the total reconnection fluxes (b). The dashed line in (b) marks 600 km s$^{-1}$ which roughly separates two types of CMEs based on their maximum velocities. The corresponding fits are also displayed in each panel.} 
    
    \label{fig:f_vr}
  \end{center}
\end{figure*}

Using flare radiation signatures and magnetograms, we have measured reconnection flux and reconnection rate. The peak reconnection rate ($\dot{\Phi}_{pk}$) spans over two orders of magnitude, from $6\times 10^{16}$ to $6\times 10^{18}$ Mx s$^{-1}$. 
In Fig.~\ref{fig:f_vr}(a), we compare $\dot{\Phi}_{pk}$ with the peak CME acceleration ($a_{pk}$). The figure shows that CMEs with larger accelerations are associated with flares with greater reconnection rates, which is expected in a theoretical work by \citet{vrvsnak2016solar}. For the 42 events, the correlation coefficient R between $\dot{\Phi}_{pk}$ and $a_{pk}$ is 0.75, and the $\dot{\Phi}_{pk}$ -- $a_{pk}$ relation can be well fitted to a power law $a_{pk}=10^{2.78} {\dot{\Phi}_{pk}}^{0.56}$, with $\dot{\Phi}_{pk}$ in uint of $10^{18}$ Mx s$^{-1}$. It is also noted that, the median (mean) peak acceleration of these 42 CMEs is 3.8 (3.5) times that of the other 18 CMEs associated with weak flares in which reconnection rate is not measured. These results provide a strong argument that flare reconnection is key to CME acceleration.

Previous studies have shown that the maximum velocities $v_{max}$ of fast CMEs are positively correlated with the total reconnection flux $\Phi_{r}$ (\citealp{qiu2005magnetic,vrvsnak2016solar,deng2017roles,toriumi2017magnetic,pal2018dependence}). Here we investigate such a relationship with the 42 events, as displayed in Fig.~\ref{fig:f_vr}(b).  We notice that, for the relatively fast CMEs with $v_{max} \gtrsim$ 600~km~s$^{-1}$ (\citealp{howard1985coronal,gopalswamy2000interplanetary}), their maximum velocity $v_{max}$ is positively correlated with the total reconnection flux $\Phi_{r}$, similar to previous findings. Here $v_{max}$ can be linearly scaled with the reconnection flux (R = 0.75) with  $v_{max}=0.86\Phi_{r}+0.73$, where $v_{max}$ has a unit of $10^3$~km~s$^{-1}$, and $\Phi_{r}$ has a unit of $10^{22}$ Mx. However, this trend breaks down for slow CMEs with maximum velocities below $\sim$600~km~s$^{-1}$.  Finally, 16 out of the 18 CMEs without reconnection flux measurements are slow CMEs and are associated with weak flares.

Our results, based on a much broader and larger event sample than previous studies, demonstrate that the CME peak acceleration is strongly related to the peak reconnection rate. Our study also confirms the positive correlation between the CME peak velocity and the total reconnection flux for fast CMEs. For slow CMEs, a significant positive correlation is still present between peak reconnection rate and peak CME acceleration, but the CME peak velocity and the flare reconnection flux are not correlated. The implication of these results will be discussed in Section 6.

\section {Temporal Analysis}

In this study, we track CMEs from the solar surface to several solar radii without data gap, with cadences as high as 75 seconds in the EUVI images and 2.5 minutes in COR1 images, and we measure the flare reconnection rate with a 24-second cadence. This is a unique dataset to perform a time series analysis of the CME evolution with respect to the progress of flare reconnection. We note that, with very few exceptions, prior studies often use the GOES soft X-ray (SXR) light curves as a proxy for flare progress to compare with CME evolution.  To relate our study with previous studies, we also examine the time evolution of CME acceleration and flare reconnection with respect to  the GOES SXR light curve.

\subsection{Time profiles of CME acceleration and flare reconnection}
Fig.~\ref{fig:f2} suggests that, on the timescale of a few minutes dictated by the observing cadence of CMEs, the CME acceleration  $a(t)$, and flare reconnection rate $\dot{\Phi}_{r}(t)$, exhibit very similar temporal profiles, rising and decaying on nearly the same timescales.
To examine the timing of the CME acceleration and flare reconnection, we conduct a time-lag analysis of $a(t)$ and $\dot{\Phi}_{r}(t)$. The analysis yields the time lag of CME acceleration with respect to flare reconnection $\Delta \tau \equiv t_a - t_r$, and the maximum correlation coefficient $R_{max\_a\_r}$ between $a(t)$ and $\dot{\Phi}_{r}(t)$ at $\Delta \tau$. Fig.~\ref{fig:f_dt} presents these results for 
the 42 events with reconnection flux measurements. $R_{max\_a\_r}$ lies within 0.68 -- 0.99, and peaks at $\sim$0.9, revealing a strong correlation between $a(t)$ and $\dot{\Phi}_{r}(t)$. 

Meanwhile, the measured time lag $\Delta \tau$ in Fig.~\ref{fig:f_dt}(b) ranges between -20 and 20 minutes, and appears to form three groups roughly centered at $-10, 0, 10$ min. Since 5-min is the synoptic  cadence of the EUVI and COR1 observations, we consider that CME acceleration leads flare reconnection for 11 events with $\Delta \tau < -5$~min,  flare reconnection and CME acceleration are simultaneous for 19 events with $-5 \le \Delta \tau \le 5$~min, and flare reconnection leads CME acceleration for the remaining 12 events of $\Delta \tau > 5$~min, respectively. The  median (and standard deviation) of $\Delta \tau$ of these three groups are $-10\pm 2.9, 1.2 \pm 2.6, 10\pm 2.8$ min, respectively, consistent with the histogram in Fig.~\ref{fig:f_dt}(b). The corresponding widths of the three peaks at $\sim$5--6 min is less than the 10-min separation between adjacent peaks, suggesting the robustness of the finding.

Examples of the three groups in Fig.~\ref{fig:f_dt}(b) are given in Fig.~\ref{fig:f_ex}. (\Rmnum{1}) Among the first group of 11 events, the C8.7 flare of SOL2011-06-21T03:27 (Fig.~\ref{fig:f_ex}(a)) corresponds to one of the strongest flares in this group. For this eruption, the CME appears to accelerate $\sim$12 min ahead of the reconnection rate. A similar event can be found in \citet{song2018acceleration}. (\Rmnum{2})  SOL2010-08-01T08:56 (Fig.~\ref{fig:f_ex}(b)) is an example among the 19 events where the $a(t)$ evolve simultaneously with $\dot{\Phi}_{r}(t)$. (\Rmnum{3}) SOL2011-12-26T20:31 (Fig.~\ref{fig:f_ex}(c)) belongs to the third group. In this event, the acceleration profile falls behind the reconnection rate by around 12 min.

As we compare the properties of magnetic reconnection and CME kinematics among these three groups, we find that all acceleration-lead events  have peak reconnection rates below 0.5$\times10^{18}$ Mx~s$^{-1}$, with the median at $0.2\times10^{18}$ Mx~s$^{-1}$, whereas the peak reconnection rate of the other two populations is higher, with the median $\sim$0.4$\times10^{18}$ Mx~s$^{-1}$, more than twice that of the acceleration-lead events. We find no significant differences among the three populations, in other properties, such as the CME initial height.

\subsection{Comparison with GOES X-ray light curves}

\begin{figure*}[ht]
  \begin{center}
        \includegraphics[viewport =  46 37 720 545, clip, width=0.95\textwidth]{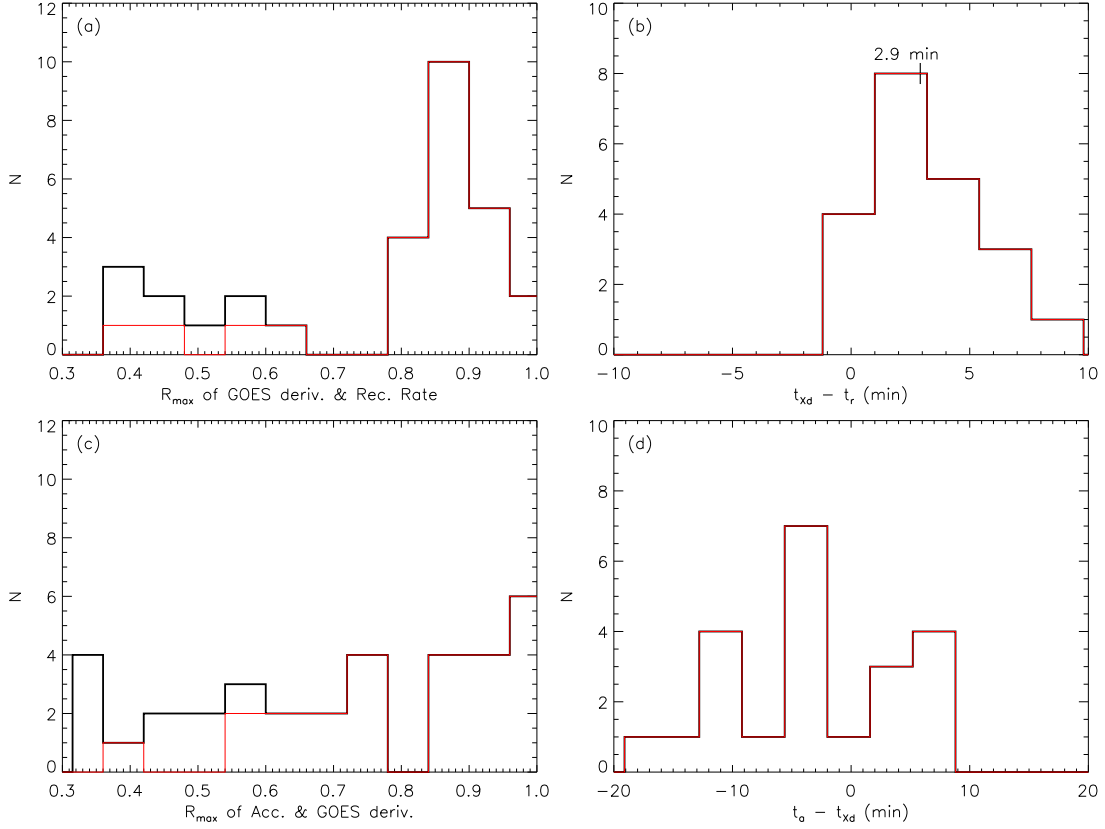}
        
    \caption{Distributions of $R_{max}$ (a) and the time differences (b) between the GOES time derivative and reconnection flux rate. Distributions of the $R_{max}$ (c) and the time differences (d) between the acceleration and GOES time derivative. Only the events with $R_{max}>0.3$ are shown in the left panels, and the events with $R_{max}>0.6$ are shown in the right panels. The distributions of events in ARs are indicated with red curves.}
    
    \label{fig:f_dt2}
  \end{center}
\end{figure*}

 The time derivative of GOES SXR profile has been often used as a proxy of the energy release, based on the Neupert effect (\citealp{neupert1968comparison,dennis1993neupert}). To allow comparison of our results with past studies, we apply the time-lag analysis between the GOES SXR time derivative, $\dot{X_r}(t)$, and the reconnection rate, $\dot{\Phi}_{r}(t)$, and CME acceleration, $a(t)$ (Fig.~\ref{fig:f_dt2}).

The relationship of $\dot{X_r}(t)$ and $\dot{\Phi}_{r}(t)$ is shown in Fig.~\ref{fig:f_dt2}(a{\&}b). Only 22 of the 42 events exhibit correlation $R_{max\_Xd\_r}$ greater than 0.6 (Fig.~\ref{fig:f_dt2}(a)). Interestingly, all 22 events originate in ARs and tend to be associated with the larger flares, which, in turn, are more consistent with the Neupert effect (\citealp{dennis1993neupert}). 
The histogram of their time differences $t_{Xd}-t_{r}$ suggests  delays of $\dot{X_r}(t)$ with respect to $\dot{\Phi}_{r}(t)$ by 1--5 min with an average of 2.9 min (Fig.~\ref{fig:f_dt2}(b)). 

Similarly, Fig.~\ref{fig:f_dt2}(c{\&}d) give the relationship of $a(t)$ to $\dot{X_r}(t)$. The same 22 events exhibit a good correlation ($R_{max\_a\_Xd}>$ 0.6). But the distribution (Fig.~\ref{fig:f_dt2}(c)) appears to be flatter than the previous two (Fig.~\ref{fig:f_dt}(a){\&}\ref{fig:f_dt2}(a)), and gives poorer correlation. Again, among the 22 events, the distribution of their time difference $t_{a}-t_{Xd}$ (Fig.~\ref{fig:f_dt2}(d)), still displays three peaks though shifted by about 3 minutes, due to the delay of $\dot{X_r}(t)$  in Fig.~\ref{fig:f_dt2}(b). 

There are several reasons for the overall poor correlation using the time derivative of the GOES SXR light curve $\dot{X_r}(t)$. After the peak SXR emission, $\dot{X_r}(t)$ becomes negative; however, reconnection energy release, reflected by the formation of new flare loops, still continues into the decay of the SXR emission, as illustrated in Fig.~\ref{fig:f2}\&\ref{fig:f_ex}. Furthermore, for weak flares, the SXR emission measured by GOES may include a significant contribution from other regions. Therefore, the magnetic reconnection rate measured in the host region is a better proxy for the energy release process in the CME-flare study. The time profile analysis in this part shows that, using the reconnection rate as the proxy, the timing of energy release is, on average, 3 minutes earlier than the timing determined using the GOES light curve.

\subsection{Start times of CME acceleration and flare reconnection}

To complement the foregoing correlation analysis on the overall progress of the CME acceleration and flare reconnection, here we investigate the timings when CME acceleration or flare reconnection begins to rise (\citealp{kahler1988filament,zhang2001temporal,temmer2008acceleration,liu2014early,wang2019evolution}). We choose the start time of CME acceleration $t_{a\_{st}}$ or that of flare reconnection rate $t_{r\_{st}}$ to be the moment when CME acceleration (or reconnection rate) has increased to 10\% of the peak (\citealp{berkebile2012relation, bein2012impulsive}). For this analysis, the time profile of the reconnection rate is smoothed with a boxcar of 5 min, similar to the time cadence of the CME observation. Fig.~\ref{fig:f_init}(a) displays a comparison of
the start times of the flare reconnection and CME acceleration for the subset of 42 CME-flare events. The time differences $\Delta \tau_{st} \equiv t_{a\_st} - t_{r\_st}$ are primarily within $\pm 25$ min, and have a median value of -1.7 min. The number of events for $\Delta \tau_{st} < -5$  min, $|\Delta \tau_{st}| \leqslant 5$ min, and $\Delta \tau_{st} > 5$ min are 16, 14, and 12, respectively. Thus the distribution is slightly asymmetric with a few more events where CME acceleration appears to rise earlier than the rise of flare reconnection. This asymmetry is consistent with but less remarkable than that using the CME acceleration and the hard X-ray emission reported by \citet{berkebile2012relation}.

Fig.~\ref{fig:f_init}(b) displays a scatter plot of $\Delta \tau_{st}$ with respect to the magnitude of GOES X-ray in 1--8 {\AA} for 35 out of the 42 events (the remainder 7 events are not included due to the presence of simultaneous strong emission from other regions or unavailability of GOES data). The scatter plot can be also divided into three groups. For 13 events with $|\Delta \tau_{st}| \leqslant 5$ min, the magnitude of the flare SXR emission spans three orders of magnitude, and the average peak reconnection rate of these events is $1.2\pm1.6\times10^{18}$ Mx~s$^{-1}$. Similarly, the 9 events with earlier flare reconnection ($\Delta \tau_{st} > 5$  min) exhibit a range of flare SXR flux, and have an average peak reconnection rate of $1.2\pm1.7\times10^{18}$ Mx~s$^{-1}$. For the remainder 13 events where CME acceleration begins earlier than flare reconnection ($\Delta \tau_{st} < -5$  min),  the amount of time lag $|\Delta \tau_{st}|$ is roughly anti-correlated with the flare SXR peak flux, \ie shorter time differences for stronger flares. The reconnection rate for this group is $0.4\pm0.4\times10^{18}$ Mx~s$^{-1}$, evidently smaller than the other two groups. 

Overall, the distribution of the time lag between the \textit{time profiles} of acceleration and reconnection rate appears to be symmetric, displaying three peaks with $\sim$10 min apart (Fig.~\ref{fig:f_dt}(b)). The distribution of the difference in the \textit{start times} is slightly asymmetric (Fig.~\ref{fig:f_init}(a)), with a few more events in which CME acceleration rises earlier than the flare reconnection. In both cases, we find that acceleration-lead events are associated with lower reconnection rates.

\section{Current-sheet Property and Energetics}
Magnetic reconnection takes place in a current sheet (CS) that forms during CME eruption. Reconnection in the CS re-configures the global magnetic field, forming flare loops and changing the CME structure, and in this course, releases free magnetic energy. The amount of released magnetic energy is equivalent to the work done by electric fields on electric currents integrated in the entire space. The reconnection rate in the CS can be described by the local reconnection electric field (\eg \citealp{Lin2000}). Approximating the flare-CME configuration by a 2D model, we can estimate the mean reconnection electric field and the electric current along the CS, using the measured reconnection rate in terms of $\dot{\Phi}_r(t)$ and the length of flare ribbons $L_R$. We can then estimate the amount of work done by the reconnection electric field in the CS, and compare it with the CME energy.

Properties of the CS can be estimated by applying the minimum-current corona \citep[MCC; ][]{longcope1996topology} to a single
current sheet following a separator of total length $L_R$.  If magnetic flux $\Phi_r$ reconnects across this separator it must
initially carry a net current with order of magnitude $I_0\sim\Phi_r/\mu_0L_R$.  The mean reconnection electric field will be
\begin{equation*}
  \langle E \rangle ~=~ |\dot{\Phi}_r(t)| / L_R ~~,
\end{equation*}
and the total magnetic energy released will be
\begin{equation*}
  W_r =\int I\, d\Phi_r ~\sim~ (\Phi_r)^2/2\mu_0 L_R ~~.
\end{equation*}

\noindent where $\mu_0$ is the permeability of free space. The MCC model addresses the build-up and release of magnetic energy in a three-dimensional corona. For the energy release process, it assumes that reconnection at the current in the CS has decreased to zero. Thus $W_r$ here corresponds to an upper limit.  

\begin{figure*}[ht]
  \begin{center}
        \includegraphics[viewport = 46 425 498 613, clip, width=0.95\textwidth]{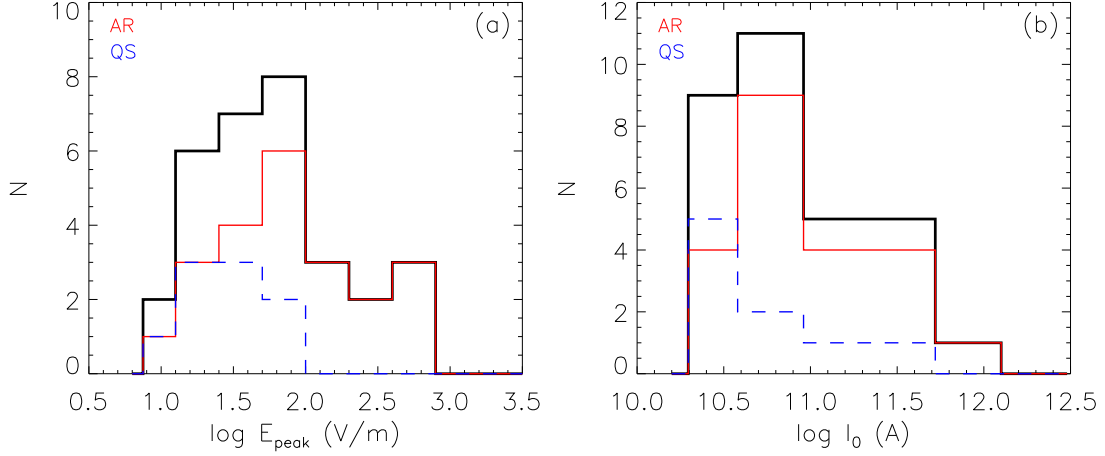}
        
    \caption{Histograms of the average electric field strength in the current sheet at the timing of peaking reconnection rates (a), and the integrated current intensity in the current sheet before the CME eruption (b).}
    
    \label{fig:f_ei}
  \end{center}
\end{figure*}

The distributions of the peak $\langle E \rangle$ and $I_0$ are shown in Fig.~\ref{fig:f_ei}(a{\&}b). $\langle E \rangle_{pk}$, determined using the peak reconnection rate, ranges 5 -- 500 V/m. The values here are typical when compared to the studies by \citet{qiu2004magnetic}, \citet{jing2005magnetic}, and \citet{song2013study}. The median value of $\langle E \rangle_{pk}$ in AR (95.5 V/m) is about 5 times as large as that in QS (20.6 V/m). Meanwhile, the pre-reconnection current intensity $I_0$, ranging between $1.3 - 72 \times10^{10}$~A, displays a stronger median value of $6.7{\times}10^{10}$~A in ARs compared to $3.0{\times}10^{10}$ A in QS. Finally, the estimated work done by the reconnection electric field, $W_r$, is in the range $10^{22-25}$ J.

\begin{figure*}[ht]
  \begin{center}
        \includegraphics[viewport = 36 232 725 615, clip, width=0.96\textwidth]{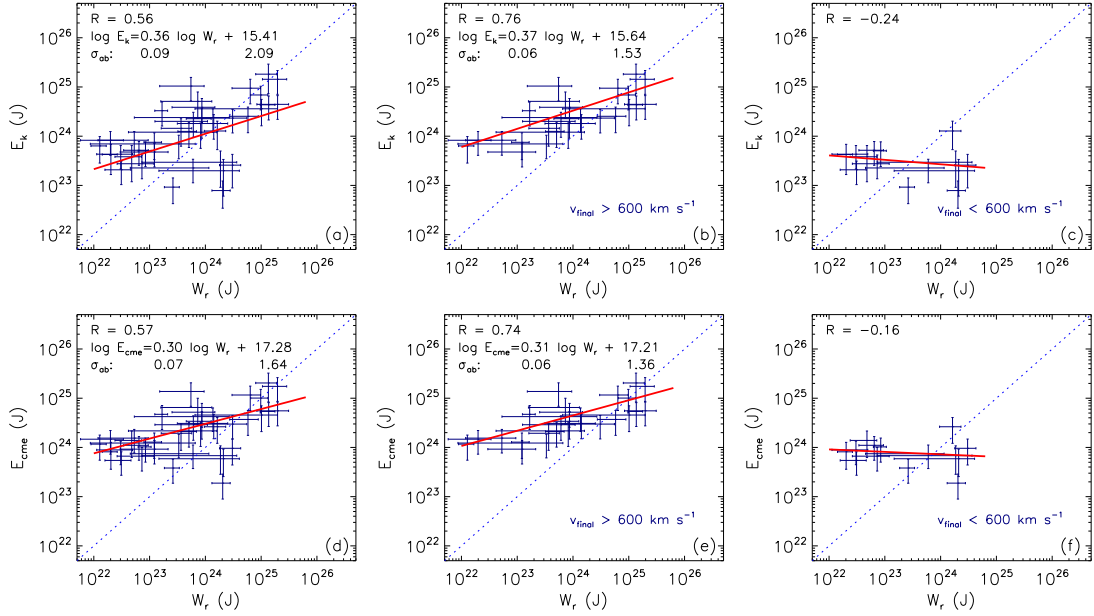}
        
    \caption{Comparisons of the CME kinetic energy $E_k$ (top panels), CME total mechanical energy ($E_{cme} = E_k+E_p$; lower panels) with the released energy $W_r$ related to the current sheet during the CME-flare events. Left panels (a\&d): Total 40 events with available measurements of both reconnection flux and CME mass. Middle panels (b\&e): 26 events with $v_{max} \gtrsim 600~km~s^{-1}$. Right panels (c\&f): 14 events with $v_{max} \lesssim 600~km~s^{-1}$. The diagonal dashed line denotes equality.}
    
    \label{fig:f12}
  \end{center}
\end{figure*}

Fig.~\ref{fig:f12} gives the relationship between $W_r$ and the CME mechanical energy ($E_{cme} \sim E_k+E_p$) for the studied CME-flare events, where $E_k$ is the CME kinematic energy, and $E_p$ is its gravitational potential energy. The CME masses, based on the theory of Thomson scattering (\citealp{billings1966guide,hayes2001deriving,howard2011coronal,vourlidas2006proper}), are derived by following the procedures described in \citet{vourlidas2010comprehensive} and making use of the COR2 data and \textit{scc\_calc\_cme\_mass.pro} in the SolarSoftware. We measure the CME mass at a time when the CME arrives at $\sim$10 \rsun. The angle from the POS is determined because of the multi-viewpoint observations here, to be the same as $\theta$. This gives  CME masses ranging from $8.7{\times}10^{14}$ to $1.9{\times}10^{16}$ g. The error of the CME masses is estimated to be a factor of 2 (\citealp{vourlidas2010comprehensive}). In this way, $E_k$ is estimated by using $E_k=m{v^2_{max}}/2$. 
We were able to measure the CME mass for 40 events. Two could not be analyzed due to a lack of  good pre-event images.

Fig.~\ref{fig:f12} shows that the CME energy ($E_k$ or $E_{cme}$) is roughly scaled with $W_r$. We first discuss $E_k$ -- $W_r$ relation (top panels in Fig.~\ref{fig:f12}). The correlation coefficient between these two quantities for the 40 events is 0.56, and $E_k$ -- $W_r$ can be fitted by a power-law function, with the power-law index of 0.36. If we only consider the relatively fast CME events with $v_{max} \gtrsim$ 600~km~s$^{-1}$, a stronger correlation (R = 0.76) is found, and $E_k = 10^{15.64} {W_r}^{0.37}$. There is no clear trend for slower CMEs with $v_{max} \lesssim$ 600~km~s$^{-1}$. Similar relations hold for the total CME energy, $E_{cme}$, and $W_r$ (lower panels in Fig.~\ref{fig:f12}). 

The diagonal dashed line in the figure marks equality and separates the energies into two regions: below this line, $E_{cme}$ $< W_r$, so $W$ may provide the necessary energy for the eruption; For events above this line, $E_{cme}$ $> W_r$, therefore $W_r$ is inadequate and the eruption requires additional energy sources. We discuss the $E_{cme}$ -- $W_r$ relationship further in Section 6.

\section{Conclusions and Discussion}
\subsection{Summary}
We conduct a statistical study of 60 CME-flare events by combining observations from different viewpoints by SDO and STEREO. The on-disk observations from the SDO provide measurements of the photospheric magnetic field and flare ribbon evolution, enabling the estimation of the magnetic reconnection flux (and reconnection rate) and the flare-ribbon geometry. The limb views from STEREO (EUVI, COR1 and COR2) yield measurements of the height of the Earth-side CMEs from 0 to 15 {\rsun}. Based on these measurements, we study the kinematics of these 60 CMEs, and examine the magnitude and time evolution of the CME acceleration with respect to the progress of magnetic
reconnection in 42 events (in which reconnection rate can be measured). We then compare the CME mechanical energy with the estimated amount of work done by the reconnection electric field in the CS. Our major findings are summarized as follows.

\begin{enumerate}
\item A strong correlation ($R_{max\_a\_r} \gtrsim 0.7$) exists between the CME acceleration and the reconnection rate (Fig.~\ref{fig:f_dt}(a)), regardless of whether CMEs are fast or slow. The CME peak acceleration shows a power-law scaling with the peak reconnection rate (Fig.~\ref{fig:f_vr}(a)). A positive correlation is also found between the maximum speed of CMEs and the total reconnection flux (Fig.~\ref{fig:f_vr}(b)), but only in relatively fast CMEs ($v_{max} \gtrsim 600~$km~s$^{-1}$), whereas no clear correlation is found for the slower CMEs.

\item The CME acceleration and flare reconnection exhibit very similar time profiles from rise to decay, and the temporal correlation between the two is much stronger than the correlation between either of them and the time derivative of the GOES SXR light curve. For the studied 42 events, the distribution of the time lag of the CME acceleration with respect to the reconnection rate ($\Delta \tau \equiv t_a - t_r$) displays a symmetric distribution with three peaks at -10.0$\pm$2.9 (11 events), 1.2$\pm$2.6 (19), and 10.0$\pm$2.8 min (12), respectively (Fig.~\ref{fig:f_dt}(b)). For a comparison, the distribution of the time lag of their start times ($\Delta \tau_{st} \equiv t_{a\_st} - t_{r\_st}$) is slightly asymmetric (Fig.~\ref{fig:f_init}(a)), with a few more events where acceleration begins earlier than the flare reconnection. On average, the peak reconnection rate of the acceleration-lead population is smaller than that of the other two populations.  
 
\item The CME total mechanical energy $E_{cme}$ displays a good correlation (R $\sim$ 0.6) with the estimated work $W_r$ done by the reconnection electric field in the CS. $E_{cme}$ -- $W_r$ can be fitted to a power-law function with the power-law index of $\sim$1/3 (Fig.~\ref{fig:f12}).    

\end{enumerate}

Past studies have shown a scaling law between CME maximum velocity and total reconnection flux, for relatively fast CMEs and strong flares (\citealp{qiu2005magnetic,deng2017roles,toriumi2017magnetic,tschernitz2018reconnection,pal2018dependence}). In this study, making use of observations from multiple views, we have analyzed a broad range of CME-flare events, \ie B to X-class flares, and both slow and fast CMEs. Our results from this extended sample demonstrate a significant correlation between CME acceleration and flare reconnection in various aspects. These results provide a strong argument that flare reconnection is key to acceleration of both fast and slow CMEs, and may dominate the acceleration of fast CMEs.

On the other hand, the finding that the maximum speed of slow CMEs is not correlated with the total reconnection flux may indicate that other physical processes play a more important role during the acceleration of these CMEs. Note that the $a_{pk}$ -- $\tau_a$ scaling indicates that, statistically speaking, an event with smaller peak acceleration (and also a smaller peak reconnection rate) tends to be accelerated for a longer period of time; it seems that, for slow CMEs, the effect of CME acceleration by flare reconnection during this extended time period is relatively weak in comparison with other mechanisms. 

Our results suggest that the significance of flare reconnection in CME acceleration is different in fast and slow CMEs. The results from the energetics and time-lag analyses bear similar indications, which will be further discussed.

\subsection{Energetics}

Our analysis reveals a good correlation between the CME mechanical energy and the estimated electrodynamic work $W_r$ done by the reconnection electric field in the CS, especially for the relatively fast CMEs (Fig.~\ref{fig:f12}). For events with $W_r \gtrsim 2\times10^{24}$~J (X and M flares; Fig.~\ref{fig:f12}(e)), $W_r$ is larger than or comparable to $E_{cme}$ and should be able to provide a significant amount of energy for the CME eruption.  This result supports the argument that flare reconnection dominates CME acceleration in fast CMEs.

However, for events with smaller flares, \ie B and C flares, $E_{cme}$ is 1 -- 2 orders of magnitude larger than $W_r$. This large ratio of $E_{cme}/W_r$ would suggest that, in these events, the work done by the reconnection electric field in the CS itself might not be enough to fuel the eruption. It is possible that reconnection also occurs elsewhere, such as the breakout type of reconnection (\citealp{Antiochos1999, Karpen2012mechanisms}). In events with very weak or little flare radiation signature, reconnection may occur at higher altitudes and produce less plasma heating, as suggested by \citet{robbrecht2009} and \citet{vourlidas2018}. Signatures of these reconnection events are either undetectable by current instrumentation or require different analysis methodologies \citep{robbrecht2010}.

Apart from reconnection, several other mechanisms can also impact the kinematics of these events. If a flux rope is present -- whether it is generated prior to or at the time of the eruption -- and carries a sizable amount of electric current, a significant amount of work can be done on this flux rope current during its eruption. Mass unloading may also help accelerate a CME (\citealp{low1996solar,klimchuk2001theory}). It is likely that these other mechanisms are more important in events characteristics of slow CMEs and weak reconnection.

The $E_{cme}$ -- $W_r$ scaling (Fig.~\ref{fig:f12}) may also shed light upon the increased association rate between CMEs and flares production rates with the rising flare intensity reported by \citet{yashiro2005visibility}. For small flares, the requirement of additional 1--2 order larger energy sources would make it very difficult to drive a CME. Thus the free energy associated with the current sheet(s) can be either released through a series of small flares without CMEs,  or wait  until additional energies become available (\eg formation of flux ropes). On the other hand, for large flares where $W_r \gtrsim E_{cme}$, they require little additional energy. Thus it would appear to be easier for the large flares to produce simultaneous CMEs. A more comprehensive evaluation of various energies, the CME energy, flare energy, and free magnetic energy including $W_r$, will be conducted in future studies.    

\subsection{Timing}
Our unique dataset with CMEs observed in the low corona and flares observed on the disk provides the advantage to examine relative
timings of the CME acceleration and flare reconnection in the early phase of the eruption. In this study, a time-lagged correlation analysis is performed to compare the temporal evolution of CME kinematics with respect to the progress of flare reconnection. The resultant time lag is determined by the slow-varying component of the time profiles, in particular, the times of the global maximum of the analyzed time series (see Fig.~\ref{fig:f_ex}). Our analysis reveals the presence of all three populations, namely, the acceleration-lead, the simultaneous, and the reconnection-lead events, and adjacent populations are separated by about 10 minutes.

These results may be discussed with respect to some eruption models. For example, in the framework of a 2D loss-of-equilibrium (LOE) model \citep{Lin2000}, \cite{reeves2006relationship} has shown that reconnection energy release rate -- the product of the reconnection electric field and magnetic field -- in the CS beneath the erupting flux rope always lags the CME acceleration, and the amount of the lag grows when either the reconnection rate (represented by the Alfv\'{e}n Mach number in the CS) or the background magnetic field decreases. The time lag analysis from this model is qualitatively consistent with the observed acceleration-lead events; but this model, and perhaps other models that drive eruption ideally, cannot explain other events in which flare reconnection leads, or evolves simultaneously with, CME acceleration. We recognize that a detailed model-observation comparison in terms of the time lag requires consideration of many other properties, such as the geometric properties and aerodynamic drag (\eg \citealp{chen2003acceleration, vrvsnak2016solar}), and is beyond the scope of the present study; nevertheless, this study has established an observational database to help construct or constrain parameter studies with numerical models.

\begin{figure*}[ht]
  \begin{center}
        \includegraphics[viewport = 46 295 720 545, clip, width=0.95\textwidth]{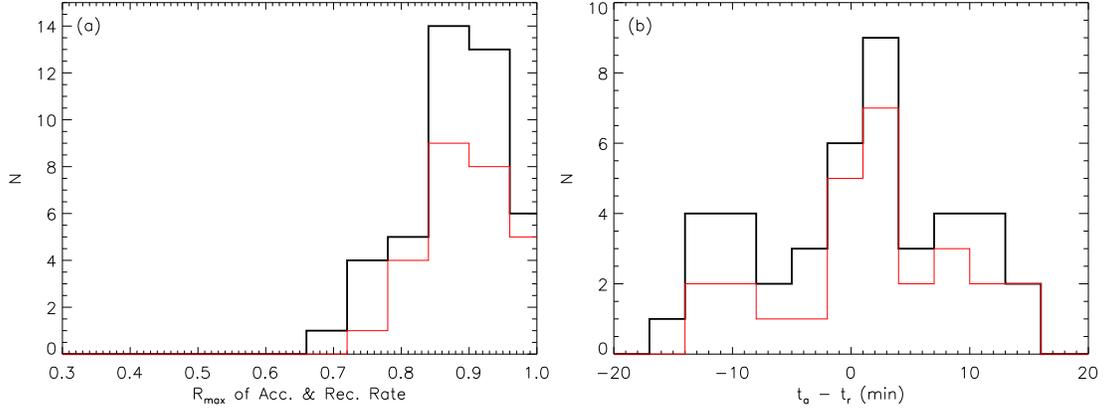}
        
    \caption{Distributions of the maximum correlation coefficient $R_{max}$ between the shifted profiles of acceleration and reconnection flux rate (a), and their time differences $t_a-t_r$ (b). A negative time difference here means that the acceleration profiles comes earlier, vice versa. The distributions of events in ARs are indicated with red curves.}
    
    \label{fig:f_dt}
  \end{center}
\end{figure*}

\begin{figure*}[ht]
   \begin{center}
         \includegraphics[viewport = 95 227 431 755, clip, width=0.325\textwidth]{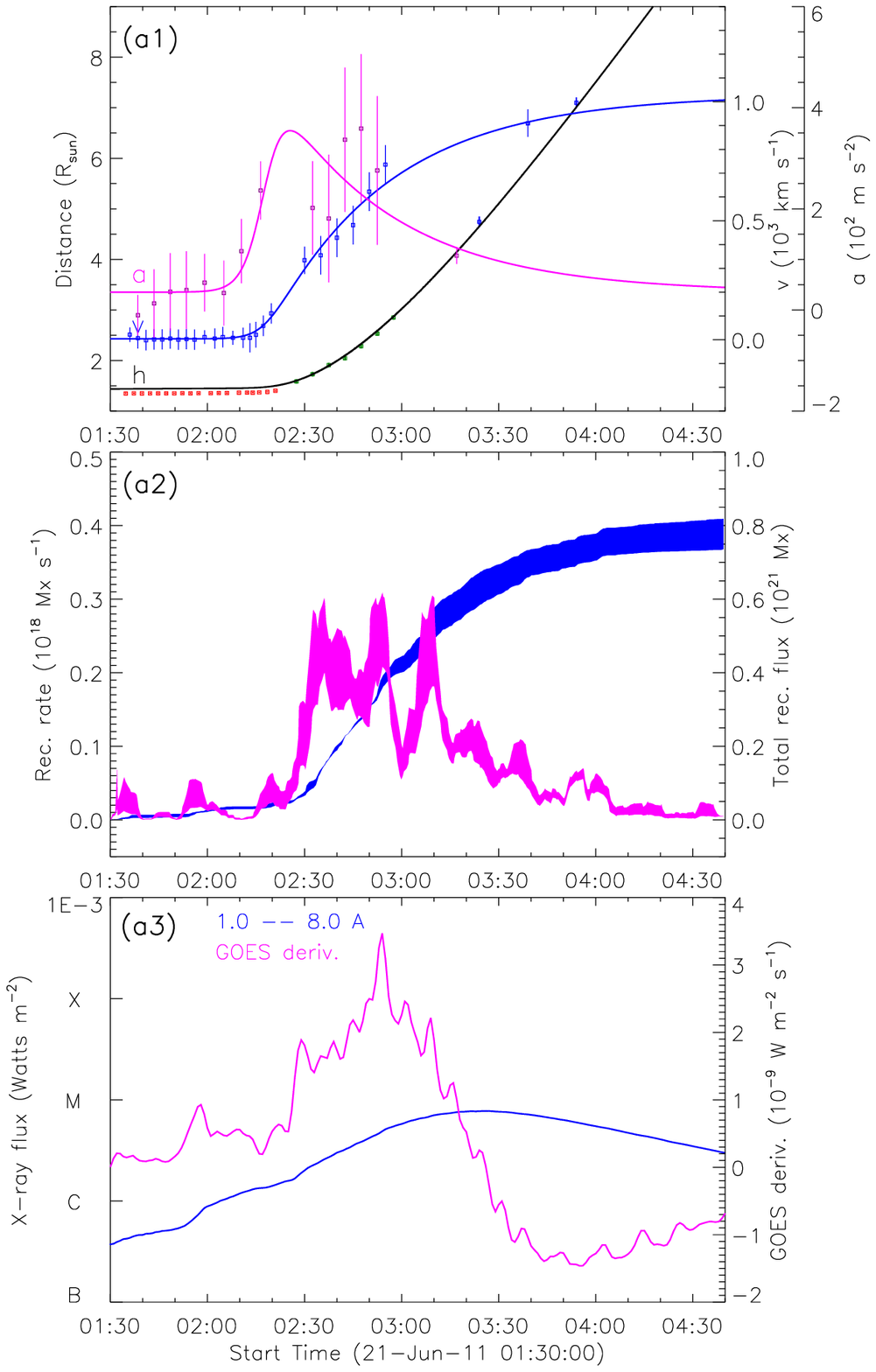}
         \includegraphics[viewport = 95 227 431 755, clip, width=0.325\textwidth]{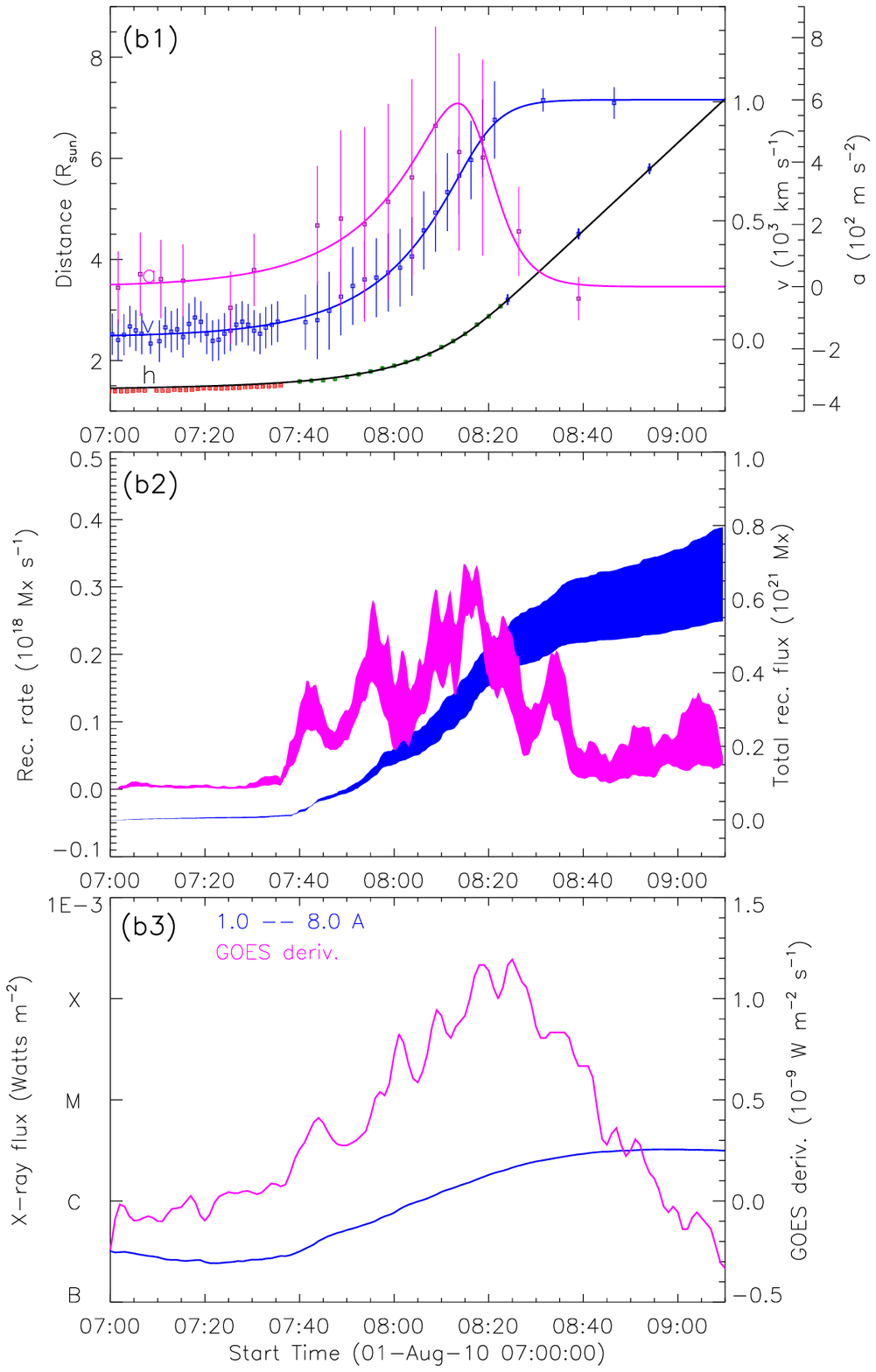}
         \includegraphics[viewport = 95 227 431 755, clip, width=0.325\textwidth]{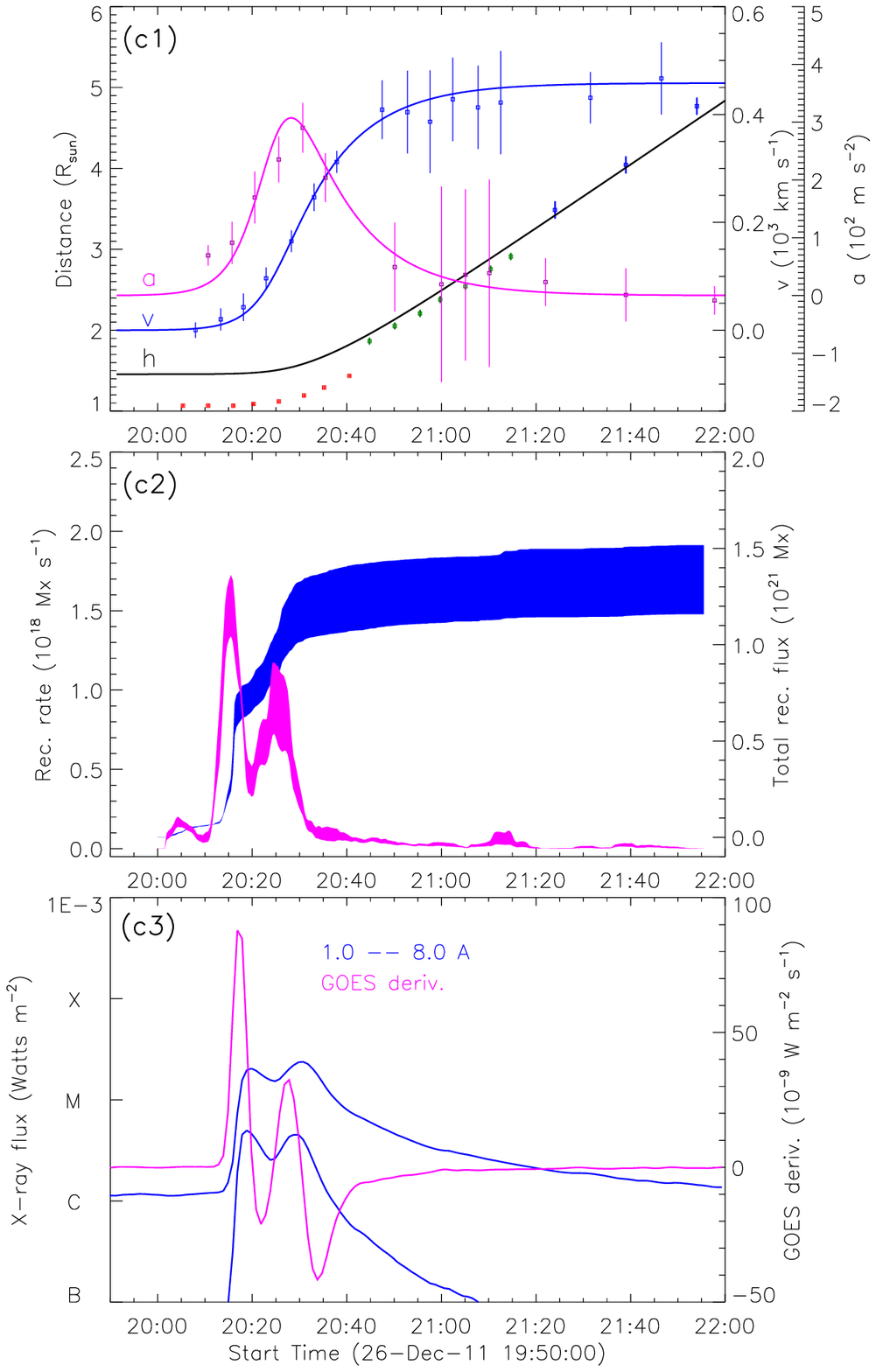}
     \caption{Three examples of the relationships between the CME kinematics, the reconnection flux, and flare X-rays, corresponding to the three peaks in Fig.~\ref{fig:f_dt}(b), using the format of Fig.~\ref{fig:f2}: the profile of the CME acceleration $a(t)$ leads the reconnection rate (a, SOL2011-06-21T03:27), simultaneously (b,  SOL2010-08-01T08:56), and $a(t)$ follows the reconnection rate (c, SOL2011-12-26T20:31). }
    
     \label{fig:f_ex}
   \end{center}
 \end{figure*}

\begin{figure*}[ht]
  \begin{center}
        \includegraphics[viewport = 132 335 640 545, clip, width=0.95\textwidth]{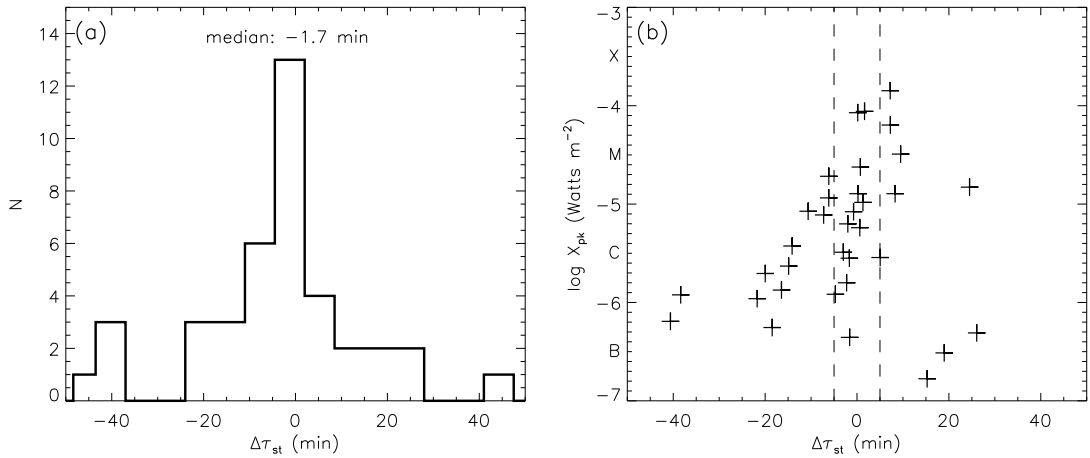}
        
    \caption{Time difference between the start times of CME acceleration and flare reconnection rate, \ie $\Delta \tau_{st} \equiv t_{a\_st} - t_{r\_st}$. Panel (a) displays a histogram of $\Delta \tau_{st}$. Panel (b) shows a scatter plot of $\Delta \tau_{st}$ vs. peaking GOES X-ray flux, with $\pm$5 min indicated by the dashed lines.}
    \label{fig:f_init}
  \end{center}
\end{figure*}

An acceleration-lead event is likely {\em driven} by an ideal instability, such as in the LOE model, yet it is still possible that the eruption is {\em triggered} by reconnection \citep{green2018origin}.  We note that the time-lagged correlation analysis (Fig.~\ref{fig:f_dt}) is not suited to answer the ``trigger" question, namely, whether the start of the fast CME motion occurs before or after the onset of flare reconnection (\eg \citealp{Karpen2012mechanisms}). We also attempted to estimate the start times of CME acceleration and flare reconnection (Fig.~\ref{fig:f_init}), using the analytical function to describe CME acceleration and much smoothed time profiles of flare reconnection. The resultant time lag distribution is similar to that of the correlation analysis, which is also likely governed by the slow-varying trend of the time profiles, and is subject to the uncertainties primarily determined by the time cadence (5 min) of CME observations. To probe the ``trigger" of the eruption in the future study, it is crucial to explore new methods, identify fast-varying features in the early stage using unsmoothed time profiles, and examine their associated signatures in the low corona prior to the eruption. This requires analysis of flare and CME observations at high temporal and spatial resolutions with the combination of all available capabilities including SDO and RHESSI (\citealp{lin2003reuven}).

\acknowledgments
{The authors thank the referee for several constructive comments. We thank Dr.~Dana~W.~Longcope for helpful discussion. The authors gratefully acknowledge the funding of NASA Heliophysics Guest Investigator (HGI) program (80NSSC18K0622). C.Z. and J.Q. also thank the funding of NASA HGI 80NSSC19K0269. This work is also supported by NSF REU program (1156011) at Montana State University. 

\facilities{\textit{SDO}, \textit{STEREO}, \textit{GOES}}}
\newpage

\begin{center}
\startlongtable
\tabletypesize{\scriptsize}
\renewcommand*{\arraystretch}{1.0}
\begin{longrotatetable}
\begin{deluxetable}{cccrrrrrrrrrrrrrr}
\tablecaption{List of 60 CME-flare events observed by STEREO and SDO.}\label{tab:table}
\tablehead{
\colhead{Event} & \colhead{STE.} & \colhead{SDO\_xy} & \colhead{AR}  & \colhead{Sep.} & \colhead{EUVI} & \colhead{h$_{00}$} & \colhead{a$_{pk}$}  & \colhead{$\tau_{a}$} & \colhead{v$_{max}$} & \colhead{rec.~wav.} & \colhead{$\Phi_{r}$} & \colhead{$\dot{\Phi}_{pk}$}& \colhead{GOES} &
\colhead{$m_{cme}$} & \colhead{$E_{p}$} \\
\colhead{$t_{a\_peak}$} & \colhead{a,b} & \colhead{arcsec} & \colhead{num}  & \colhead{deg} & \colhead{\AA} & \colhead{\rsun} & \colhead{$m~s^{-2}$}  & \colhead{min} & \colhead{$km~s^{-1}$} & \colhead{\AA} & \colhead{$10^{21} Mx$} & \colhead{$10^{18} Mx~s^{-1}$} & \colhead{class} &
\colhead{$10^{15} g$} & \colhead{$10^{23} J$} 
} 
\startdata
01		20100801 08:13	&	a	&	[-480,~190]	&	11092	&	78	&	171	&	0.51	&	588	&	23.1	&	1008	&	1600	&	1.4$\pm$0.70	&	0.31$\pm$ 0.04	&	C3.2	&	4.0	&	6.5	\\
02		20100807 18:08	&	a	&	[-530,~170]	&	11093	&	79	&	171	&	0.22	&	931	&	11.2	&	747	&	1600	&	1.6$\pm$0.39	&	1.28$\pm$ 0.12	&	M1.0	&	6.7	&	11.3	\\
03		20100911 01:33	&	a	&	[-400,~450]	&		&	82	&	171	&	0.58	&	82	&	52.8	&	326	&	304	&	0.4$\pm$0.10	&	0.07$\pm$ 0.01	&	B3.1	&	6.1	&	10.1	\\
04		20101130 18:40	&	a	&	[-600,~230]	&		&	85	&	195	&	0.53	&	131	&	48.2	&	490	&	304	&	0.2$\pm$0.04	&	0.07$\pm$ 0.01	&	B1.7	&	2.4	&	3.9	\\
05		20101214 15:17	&	a	&	[~700,~400]	&	11133	&	85	&	195	&	1.36	&	241	&	42.9	&	771	&		&		&		&	C2.3	&		&		\\
06		20110216 02:17	&	b	&	[-400,~500]	&		&	94	&	195	&	0.66	&	63	&	57.5	&	266	&	304	&	0.4$\pm$0.13	&	0.06$\pm$ 0.01	&		&	2.4	&	3.9	\\
07		20110621 02:30	&	a	&	[~100,~240]	&	11236	&	96	&	171	&	0.47	&	318	&	41.6	&	1001	&	1600	&	2.5$\pm$1.76	&	0.27$\pm$ 0.14	&	C7.8	&	5.8	&	9.5	\\
08		20110709 00:20	&	b	&	[-350,-330]	&	11247	&	92	&	171	&	0.45	&	275	&	33.8	&	690	&	304	&	0.8$\pm$0.14	&	0.21$\pm$ 0.03	&	B4.9	&	3.2	&	5.3	\\
09		20110802 06:19	&	a	&	[~240,~160]	&	11261	&	100	&	171	&	0.49	&	521	&	20.6	&	791	&	1600	&	3.6$\pm$1.02	&	1.12$\pm$ 0.08	&	M1.5	&	7.1	&	11.7	\\
10		20110910 02:59	&	a	&	[~300,~450]	&		&	103	&	171	&	0.80	&	93	&	95.7	&	645	&	304	&	0.2$\pm$0.10	&	0.08$\pm$ 0.02	&	B5.5	&	3.9	&	6.4	\\
11		20110913 22:58	&	a	&	[~220,~220]	&		&	103	&	171	&	0.61	&	152	&	29.3	&	342	&	304	&	2.2$\pm$0.48	&	0.47$\pm$ 0.05	&		&	4.3	&	7.0	\\
12		20111001 09:48	&	a	&	[~~90,~~50]	&	11305	&	104	&	195	&	0.19	&	405	&	17.9	&	533	&	1600	&	1.9$\pm$0.41	&	0.78$\pm$ 0.02	&	M1.3	&	9.0	&	13.6	\\
13		20111022 01:40	&	a	&	[~500,~400]	&		&	105	&	195	&	1.27	&	66	&	136.2	&	634	&		&	 	&		&		&		&		\\
14		20111107 23:45	&	a	&	[~450,~450]	&		&	106	&	195	&	0.61	&	148	&	49.6	&	542	&	304	&	0.3$\pm$0.11	&	0.08$\pm$ 0.02	&		&	2.9	&	4.6	\\
15		20111109 08:30	&	a	&	[-250,-500]	&		&	106	&	195	&	0.79	&	67	&	74.4	&	380	&		&	 	&		&		&		&		\\
16		20111109 13:08	&	b	&	[-500,~320]	&	11342	&	103	&	195	&	0.90	&	568	&	15.9	&	658	&	1600	&	2.1$\pm$1.00	&	1.11$\pm$ 0.32	&	M1.2	&	7.7	&	12.7	\\
17		20111114 20:08	&	a	&	[~480,-450]	&		&	106	&	171	&	0.55	&	193	&	40.9	&	558	&		&	 	&		&		&		&		\\
18		20111117 16:10	&	a	&	[~350,-450]	&		&	106	&	195	&	0.65	&	80	&	37.2	&	246	&		&	 	&		&		&		&		\\
19		20111126 07:02	&	a	&	[~600,~250]	&	11353	&	106	&	195	&	0.77	&	474	&	23.5	&	804	&	304	&	0.9$\pm$0.11	&	0.35$\pm$ 0.04	&	C1.2	&	8.9	&	14.1	\\
20		20111225 00:20	&	a	&	[~200,~400]	&		&	107	&	195	&	0.31	&	338	&	21.6	&	521	&	304	&	0.6$\pm$0.08	&	0.30$\pm$ 0.03	&	C1.1	&	3.4	&	5.2	\\
21		20111226 11:28	&	b	&	[~~20,~350]	&	11384	&	110	&	195	&	0.65	&	733	&	15.1	&	788	&	1600	&	1.2$\pm$0.58	&	0.38$\pm$ 0.12	&	C5.8	&	5.4	&	8.9	\\
22		20111226 20:28	&	a	&	[~600,-360]	&	11387	&	107	&	195	&	0.61	&	307	&	20.5	&	458	&	1600	&	1.6$\pm$0.30	&	2.31$\pm$ 0.30	&	M2.4	&	0.7	&	1.1	\\
23		20120119 10:45	&	a	&	[~600,~600]	&		&	108	&	195	&	1.11	&	31	&	154.4	&	309	&		&	 	&		&		&		&		\\
24		20120119 14:47	&	b	&	[-350,~600]	&	11402	&	113	&	195	&	1.09	&	316	&	47.2	&	1122	&	1600	&	2.8$\pm$0.80	&	0.65$\pm$ 0.10	&	M3.2	&	13.0	&	21.8	\\
25		20120123 03:42	&	b	&	[~300,~580]	&	11402	&	114	&	171	&	1.60	&	1000	&	10.5	&	763	&	1600	&	4.9$\pm$0.86	&	2.50$\pm$ 0.01	&	M8.8	&	19.4	&	31.9	\\
26		20120210 19:51	&	b	&	[-200,~550]	&		&	116	&	195	&	1.24	&	128	&	59.0	&	534	&	304	&	0.4$\pm$0.19	&	0.10$\pm$ 0.04	&		&	4.3	&	7.2	\\
27		20120224 03:16	&	b	&	[-450,~500]	&		&	117	&	195	&	0.18	&	168	&	38.1	&	480	&		&	 	&		&		&		&		\\
28		20120309 03:46	&	b	&	[~~50,~400]	&	11429	&	118	&	171	&	0.66	&	1171	&	13.5	&	1149	&	1600	&	5.8$\pm$0.88	&	4.96$\pm$ 0.52	&	M6.4	&	6.6	&	10.5	\\
29		20120310 17:31	&	a	&	[~350,~400]	&	11429	&	110	&	171	&	0.11	&	1234	&	13.1	&	1158	&	1600	&	6.8$\pm$1.44	&	5.88$\pm$ 0.47	&	M8.5	&	13.6	&	21.9	\\
30		20120405 21:01	&	a	&	[~480,~370]	&	11450	&	111	&	171	&	0.63	&	445	&	20.5	&	648	&	1600	&	1.1$\pm$0.82	&	0.11$\pm$ 0.04	&	C1.6	&	10.8	&	18.4	\\
31		20120419 15:46	&	b	&	[-620,-350]	&		&	119	&	171	&	0.13	&	75	&	126.5	&	677	&		&	 	&		&		&		&		\\
32		20120429 08:41	&	b	&	[-650,~400]	&		&	118	&	195	&	0.48	&	84	&	76.5	&	539	&		&	 	&		&	B5.6	&		&		\\
33		20120507 00:58	&	a	&	[~500,-300]	&	11471	&	114	&	195	&	0.51	&	170	&	50.3	&	629	&	1600	&	0.9$\pm$0.64	&	0.19$\pm$ 0.03	&	C2.0	&	6.1	&	9.6	\\
34		20120508 09:33	&	a	&	[~~50,~300]	&		&	114	&	195	&	0.40	&	345	&	16.4	&	412	&	1600	&	0.7$\pm$0.56	&	0.15$\pm$ 0.06	&		&	2.6	&	3.9	\\
35		20120603 11:58	&	b	&	[-620,-300]	&		&	117	&	195	&	0.87	&	104	&	51.2	&	385	&	1600	&	0.5$\pm$0.37	&	0.10$\pm$ 0.01	&		&	3.1	&	5.2	\\
36		20120614 13:49	&	b	&	[-100,-300]	&	11504	&	116	&	171	&	0.55	&	499	&	31.8	&	1140	&	1600	&	3.4$\pm$1.08	&	1.13$\pm$ 0.20	&	M1.9	&	6.6	&	11.0	\\
37		20120712 16:22	&	a	&	[~~20,-300]	&	11520	&	120	&	171	&	0.67	&	552	&	27.7	&	1188	&	1600	&	6.9$\pm$1.70	&	2.88$\pm$ 0.11	&	X1.4	&	18.5	&	30.7	\\
38		20120814 02:46	&	a	&	[~300,-320]	&	11542	&	123	&	195	&	0.33	&	75	&	59.7	&	334	&	304	&	0.7$\pm$0.11	&	0.21$\pm$ 0.01	&	C1.2	&	1.6	&	2.9	\\
39		20120817 23:03	&	a	&	[~650,~380]	&		&	123	&	195	&	0.58	&	298	&	19.7	&	470	&		&	 	&		&		&		&		\\
40		20120831 19:41	&	b	&	[-600,-400]	&	11562	&	116	&	171	&	0.18	&	814	&	16.3	&	949	&	1600	&	1.4$\pm$0.58	&	0.38$\pm$ 0.19	&	C8.5	&	20.2	&	32.7	\\
41		20120927 23:40	&	a	&	[~490,~~80]	&	11577	&	126	&	195	&	0.83	&	892	&	16.3	&	1052	&	1600	&	1.1$\pm$0.51	&	0.43$\pm$ 0.13	&	C3.8	&	8.5	&	13.8	\\
42		20120929 08:45	&	a	&	[~600,~~40]	&		&	126	&	195	&	0.60	&	168	&	34.6	&	426	&		&	 	&		&		&		&		\\
43		20121023 08:07	&	a	&	[~380,~180]	&	11593	&	127	&	195	&	0.77	&	173	&	29.4	&	372	&	1600	&	1.4$\pm$0.98	&	0.32$\pm$ 0.11	&	C2.9	&	2.5	&	3.9	\\
44		20121120 11:55	&	a	&	[~350,~200]	&		&	128	&	195	&	0.66	&	181	&	51.5	&	660	&	304	&	1.0$\pm$0.20	&	0.32$\pm$ 0.05	&		&	5.2	&	8.5	\\
45		20121122 08:57	&	b	&	[-280,-380]	&		&	125	&	195	&	0.80	&	91	&	75.0	&	480	&	304	&	0.4$\pm$0.09	&	0.11$\pm$ 0.02	&	B6.4	&	3.5	&	5.9	\\
46		20121126 04:27	&	a	&	[~300,-480]	&		&	128	&	195	&	0.58	&	129	&	55.0	&	427	&	304	&	0.4$\pm$0.10	&	0.12$\pm$ 0.04	&	B7.3	&	2.1	&	3.4	\\
47		20130206 02:46	&	b	&	[-500,~440]	&		&	137	&	195	&	0.63	&	137	&	60.6	&	618	&	1600	&	0.5$\pm$0.40	&	0.10$\pm$ 0.07	&	C1.3	&	3.3	&	5.3	\\
48		20130212 22:58	&	a	&	[~720,-460]	&		&	130	&	171	&	0.83	&	310	&	38.6	&	854	&		&	 	&		&	B5.9	&		&		\\
49		20130301 18:48	&	a	&	[~620,-180]	&		&	131	&	195	&	0.46	&	68	&	57.7	&	364	&		&	 	&		&	B6.7	&		&		\\
50		20130316 14:15	&	a	&	[~750,~450]	&		&	132	&	195	&	0.60	&	244	&	36.2	&	587	&		&	 	&		&	B7.5	&		&		\\
51		20130404 21:22	&	a	&	[~600,-250]	&		&	133	&	195	&	0.50	&	131	&	42.7	&	400	&	304	&	0.1$\pm$0.03	&	0.08$\pm$ 0.01	&	B5.6	&		&		\\
52		20130421 20:11	&	a	&	[~750,-250]	&	11723	&	134	&	195	&	0.45	&	252	&	30.9	&	578	&	1600	&	0.4$\pm$0.17	&	0.24$\pm$ 0.02	&	C2.8	&	2.6	&	4.5	\\
53		20130511 23:42	&	a	&	[~500,~500]	&		&	136	&	195	&	0.45	&	96	&	70.6	&	493	&		&	 	&		&		&		&		\\
54		20130806 01:44	&	b	&	[-350,~320]	&		&	138	&	171	&	0.31	&	996	&	8.3	&	630	&	304	&	0.2$\pm$0.03	&	0.22$\pm$ 0.03	&	B4.4	&	3.2	&	5.3	\\
55		20130817 00:46	&	b	&	[-350,~450]	&		&	138	&	195	&	0.71	&	48	&	81.9	&	316	&		&	 	&		&	B4.6	&		&		\\
56		20130829 05:43	&	a	&	[~400,-550]	&		&	145	&	195	&	0.93	&	68	&	119.9	&	621	&		&	 	&		&		&		&		\\
57		20130830 02:20	&	b	&	[-650,~160]	&	11836	&	138	&	195	&	0.98	&	662	&	21.1	&	989	&	1600	&	1.4$\pm$0.59	&	0.49$\pm$ 0.07	&	C8.4	&	7.9	&	13.4	\\
58		20131011 12:16	&	b	&	[-570,-470]	&	11865	&	140	&	195	&	0.15	&	471	&	16.3	&	592	&	1600	&	0.6$\pm$0.33	&	0.37$\pm$ 0.06	&	C6.3	&		&		\\
59		20131026 12:41	&	a	&	[~600,-600]	&		&	148	&	195	&	0.86	&	95	&	84.1	&	590	&		&	 	&		&		&		&		\\
60		20131207 07:21	&	a	&	[~720,-250]	&	11909	&	150	&	195	&	0.16	&	1246	&	11.5	&	1016	&	1600	&	0.7$\pm$0.21	&	0.77$\pm$ 0.13	&	M1.3	&	2.7	&	4.9	\\
\enddata
\end{deluxetable}
\end{longrotatetable}
\end{center}

\acknowledgments

\bibliographystyle{aasjournal} 

\end{document}